\documentclass[twocolumn,reprint,prd,showkeys,superscriptaddress,nofootinbib,preprintnumbers]{revtex4-1}
\usepackage{graphicx}
\usepackage{amsmath}
\usepackage{subfig}
\usepackage{caption}
\usepackage{amssymb}
\usepackage{color}
\usepackage{xspace}
\usepackage{soul}
\usepackage{aas_macros}

\usepackage[tight,nice]{units}

\newcommand{\simgt}%
{\,\hbox{\lower0.6ex\hbox{$\sim$}\llap{\raise0.6ex\hbox{$>$}}}\,}

\newcommand*{\sci} [2] {\ensuremath{{#1} \times 10^{#2}}}

\usepackage{epstopdf}

\usepackage{hyperref}

\newcommand*{\avg} [1] {\ensuremath{\langle #1 \rangle}}
\renewcommand{\vec}[1]{\boldsymbol{#1}} 
\newcommand{\Nn}{\ensuremath{N_\text{neigh}}\xspace}
\newcommand{\np}{\ensuremath{n_\text{p}}\xspace}
\newcommand{\riso}{\ensuremath{r_\text{iso}}\xspace}
\newcommand*{\ie} {\emph{i.\,e.}\xspace}

\newcommand*{\mpch} [1] {\ensuremath{\unit[#1]{h^{-1}Mpc}}}
\newcommand*{\msun} {\ensuremath{{M_{\odot}}}}
\newcommand*{\msunh} [2] {\ensuremath{\unit[\sci{#1}{#2}]{h^{-1}M_{\odot}}}}
\newcommand{\rmag}{\ensuremath{r\text{-mag}}\xspace}
\newcommand{\rband}{\ensuremath{r\text{-band}}\xspace}
\newcommand{\smass}{\ensuremath{m_{\star}}\xspace}
\newcommand*{\kms} [1] {\ensuremath{\unit[#1]{km/s}}}
\newcommand*{\lcdm} {$\Lambda$CDM\xspace}
\newcommand{\rperp}{\ensuremath{r_\perp}\xspace}

\newcommand*{\tref} [1] {Table \ref{#1}\xspace}

\newcommand*{\sref} [1] {Sec.\ \ref{#1}\xspace}

\newcommand*{\fref} [1] {Fig.\ \ref{#1}\xspace}

\newcommand*{\cref} [1] {Chapter~\ref{#1}\xspace}

\begin{document}

\title{Using galaxy pairs as cosmological tracers}

\author{Alicia \surname{Bueno Belloso}}
\email{alicia.bueno@uam.es}
\affiliation{Institute of Cosmology and Gravitation, University of Portsmouth, Dennis Sciama Building, Portsmouth P01 3FX, UK.}
\affiliation{Instituto de F\'isica Te\'orica (UAM/CSIC), Universidad Aut\'onoma de Madrid, Cantoblanco 28049 Madrid, Spain.}

\author{Guido W. \surname{Pettinari}}
\affiliation{Institute of Cosmology and Gravitation, University of Portsmouth, Dennis Sciama Building, Portsmouth P01 3FX, UK.}

\author{Nikolai \surname{Meures}}
\affiliation{Institute of Cosmology and Gravitation, University of Portsmouth, Dennis Sciama Building, Portsmouth P01 3FX, UK.}

\author{Will J. \surname{Percival}}
\affiliation{Institute of Cosmology and Gravitation, University of Portsmouth, Dennis Sciama Building, Portsmouth P01 3FX, UK.}

\date{\today}
\preprint{Preprint number: IFT-UAM/CSIC-12-33}

\begin{abstract}

The Alcock-Paczynski (AP) effect uses the fact that, when analyzed with the correct geometry, we should observe structure that is statistically isotropic in the Universe. For structure undergoing cosmological expansion with the background, this constrains the product of the Hubble parameter and the angular diameter distance. However, the expansion of the Universe is inhomogeneous and local curvature depends on density.  We argue that this distorts the AP effect on small scales.  After analyzing the dynamics of galaxy pairs in the Millennium simulation \cite{springel:2005a}, we find an interplay between peculiar velocities, galaxy properties and local density that affects how pairs trace cosmological expansion. 
We find that only low mass, isolated galaxy pairs trace the average expansion with a minimum ``correction" for peculiar velocities. Other pairs require larger, more cosmology and redshift dependent peculiar velocity corrections and, in the small-separation limit of being bound in a collapsed system, do not carry cosmological information.

\end{abstract}


\maketitle

\section{Introduction}

One of the main efforts of modern cosmology is to determine what is responsible for the observed accelerated expansion of the universe \cite{riess:1998a, perlmutter:1999a}. Alcock and Paczynski \cite{alcock:1979a} proposed a cosmological test (hereafter denoted AP) based on the assumption of statistical isotropy around any comoving location. For regions of space-time that expand with the background, observed angles and redshifts can be translated into proper distances using the angular diameter distance $d_A(z)$ and the reciprocal of the Hubble parameter $H(z)$. Requiring isotropy in proper distance, after translating from angle and redshift measurements, leads to measurements of the product $d_A(z)H(z)$.

Because radial information comes from redshifts, AP measurements are traditionally limited by peculiar velocities, also known as comoving velocities \cite{ballinger:1996a, simpson:2009a}. These add to expansion-driven redshifts, leading to apparent anisotropic clustering if redshifts are assumed to be completely cosmological in origin, even if the correct $d_A(z)$ and $H(z)$ are used to analyze redshifts. These redshift-space distortions (hereafter RSD) are degenerate with the AP effect, removing signal \cite{ballinger:1996a, simpson:2009a}, unless assumptions are made such as the Universe following a FLRW metric \cite{samushia:2011a}. In fact, it is simply standard convention that makes us split redshift into cosmological and peculiar velocity components: considering that pairs of galaxies move due to local space-time curvature shows that the expansion rate and the RSD component can be strongly correlated. In the extreme case of bound systems, for example, the combined pairwise velocity is not dependent on background evolution, i.e.\ the expansion-driven redshift difference across a pair is exactly canceled by the RSD signal (see Appendix \ref{app:bound_systems}).

Marinoni and Buzzi \cite{marinoni:2010a} recently proposed a method to derive cosmological constraints from pairs of galaxies for which peculiar velocities can be modeled. They provided a fitting formula for the observed distribution of velocities, which can then be used to help break the AP-RSD degeneracy. They assume the normalization of the galaxy velocity distribution to be redshift and cosmology independent, whereas a more recent work by \citet{jennings:2012a} questioned this statement using N-body simulations. In the work presented here we investigate this further, considering how well pairs of galaxies, selected using different properties, trace the cosmological expansion. 

We use the Millennium simulation \cite{springel:2005a} to test how the pairwise velocity of galaxy pairs may contain information about the background expansion of the Universe. We argue that the local density in which the pairs are found may affect the amount of information these pairs carry on cosmology, because each patch of the Universe expands in a way that depends on the local density. Our analysis suggests that selecting isolated pairs, as considered by Marinoni and Buzzi, can result in average pairwise velocities more in line with the Hubble expansion, i.e.\ they need smaller, less cosmology dependent peculiar velocity corrections. We also find a better match if low-mass tracers are used.

The layout of this paper is as follows: in \sref{sec:AP} we briefly review the AP effect. In \sref{sec:millennium} we describe the Millennium simulation and the two semi-analytic models used: \citet{guo:2011a} and \citet{font:2008a}. In \sref{sec:all_pairs} we present and discuss the results we obtained by analyzing all galaxy pairs regardless of their local density, while in \sref{sec:isolated_pairs} we consider only isolated pairs. The effects of varying galaxy properties are studied in \sref{sec:galaxy_properties}. In \sref{sec:comparison_catalogs} we compare the results from the two different semi-analytic galaxy formation models used. We then conclude in \sref{sec:conclusions}.

\section{The Alcock Paczynski effect}  \label{sec:AP}

Consider a distribution of particles expanding with the
Hubble flow, in the redshift interval ($z-\Delta z/2,
z+\Delta z/2$) and subtended by an angle $\Delta\theta$. Assuming a
FLRW cosmology, the proper size of the object perpendicular to our line of
sight is given by
\begin{equation}
   d_1=d_A(z)\Delta\theta, 
\end{equation} 
where $d_A(z)$ is the angular diameter distance to the object. The size of the object parallel to the line of sight is given by
\begin{equation} \label{eq:d2}
  d_2=\frac{\Delta z}{(1+z)H(z)},
\end{equation}
where $\Delta z$ is the difference in the redshift of objects closest
and furthest away from the observer and $H(z)$ is the Hubble parameter
at the central redshift of the distribution.

Assuming that the collection of particles statistically does not have a preferred direction with respect to one line of sight, then
$\langle d_1\rangle =\langle d_2 \rangle$, allowing a statistical cosmological measurement \cite{alcock:1979a}, from a sufficient number of pairs, of
\begin{equation}
  H(z)d_A(z) = \frac{\Delta z}{(1+z)\Delta\theta}.
\end{equation}
Note that $\Delta z$, $z$ and $\Delta \theta$ are all directly
observable quantities. The AP effect, as described above assumes that $\Delta z$ as measured only depends on the cosmological expansion. In fact, the relative  velocity of pairs of particles depends on the local curvature of space, so this is not necessarily a good approximation.

\section{The Millennium simulation and semi-analytic galaxy models} \label{sec:millennium}

In order to quantify how the dynamics of galaxy pairs may be affected by factors like redshift, isolation radius, mass of the halo etc., we have considered a population of galaxies from  the Millennium simulation \cite{springel:2005a}. This traces the evolution of 2160$^3$ dark matter particles of mass $1.18\times 10^9\ \text{M}_\odot$ from redshift 127 to the present day inside a periodic box of side \mpch{500}. The simulation assumes a \lcdm cosmology with parameters based on a combined analysis of the 2dFGRS \cite{colless:2001a} and the first-year WMAP data \cite{spergel:2003a}. The parameters are $\Omega_m=0.25,\ \Omega_b=0.045,\ \Omega_\Lambda=0.75,\ n=1,\ \sigma_8=0.9$ and $H_0=73$ km s$^{-1}$ Mpc$^{-1}$.

Data on dark matter particles were stored at 64 different times. At each output time, the post-processing pipeline produced a friends-of-friends (FOF) catalog by linking particles with a separation less than 0.2 of the mean value. The SUBFIND algorithm \cite{springel:2001a} was then applied to each FOF group to identify all the substructures. The merger trees, vital for galaxy formation modeling, were then constructed by linking each subhalo found in a given ``snapshot" to one and only one descendant in the subsequent output time-slice.

We used the data from two semi-analytic models of galaxy formation based on the Millennium simulation: \citet{guo:2011a} and \citet{font:2008a}. Most of our analysis uses the semi-analytic model developed by \citet{guo:2011a} which is based on the growth of and merging of the population of subhaloes. Within this catalog, each FOF group hosts a central galaxy which sits in the minimum of the potential of the main subhalo. Other galaxies associated to the same FOF group may sit at the potential minima of smaller subhaloes or may no longer correspond to a resolved dark matter substructure, the latter being know as `orphans'. The last two collectives of galaxies are referred to as satellites, although in \citet{guo:2011a}, the physical processes affecting satellite galaxies only begin to differ from those affecting central galaxies when a satellite first enters within the virial radius of the larger system. This is the radius of the largest sphere with its center at the center of the FOF group and a mean overdensity exceeding 200 times the critical value.

For our analysis, we have varied several parameters from the galaxy catalogs, namely redshift, mass of the subhalo that hosts the galaxy, stellar mass and \rband rest-frame magnitude.  Unless differently stated, the redshift shown in our plots corresponds to $z=0.989$.

\section{All galaxy pairs} \label{sec:all_pairs}
In this section we study the average pairwise velocity of galaxies regardless of local density and galaxy properties.  We shall compare our findings with predictions from linear theory to examine general trends, and test the possibility of using randomly selected galaxy pairs to trace cosmological expansion.

\subsection{Method}\label{sec:method_allpairs}

For each galaxy pair, we compute the comoving separation $ d $, the pairwise velocity $ v_{12} $ and its square $ v_{12}^2 $.  We define the pairwise velocity as:
\begin{equation}
\label{eq:pairwise_velocity_definition}
  v_{12} = \frac{\operatorname{d}d}{\operatorname{d}t} = \frac{(\vec{v_1} - \vec{v_2}) \, \cdot \, (\vec{x_1} - \vec{x_2})}{d} \;,
\end{equation}
where $ t $ is cosmic time, and $\vec{x}$ and $\vec{v}$ are galaxy positions and velocities.  Note that, following the Millennium simulation, we work in coordinates that are comoving with the Hubble flow, hence, $ v_{12} $ represents the peculiar, non-Hubble component of the pairwise velocity.  In the plots that follow, we shall always show the $ - H(z)\,d $ curve and denote it as ``static solution".  We shall also highlight the zero line, which in these plots represents the Hubble flow, and denote it as ``comoving solution".  Any data point above the comoving solution represents pairs where the two galaxies are receding from each other faster than the Hubble expansion, while below they are moving towards each other in comoving coordinates.

We group pairs in bins according to their separation $ d $ and, for each bin, compute the average $ \avg{v_{12}} $ and the variance $ \text{var}(v_{12}) = \avg{v_{12}^2} - \avg{v_{12}}^2 $ of the pairwise velocity.  Our definition of expectation value is:
\begin{equation}
\label{eq:average_definition}
  \avg{v^n} \equiv \frac{1}{N_\text{pairs}} \, \sum\limits_{i=1}^{N_\text{pairs}} (v_i)^n \;,
\end{equation}
where $ N_\text{pairs} $ is the number of pairs in the separation bin.  In all the plots in this paper, we shall always show $ \avg{v_{12}} $ as a function of pair separation, with the error bars for each bin taken as $ \sqrt{\text{var}(v_{12}) / N_\text{pairs}} $.

By default, we employ a logarithmic binning in galaxy separation.  In order to better visualize the data, however, in some plots we combine underpopulated bins together so that each bin represents at least a minimum number of galaxy pairs, $ N_\text{min}$.  In this section we use $ N_\text{min} \geq 1000 $ while, due the poor statistics, we shall employ $ N_\text{min} \geq 2 $ for some of the ``isolated" plots in Sections \ref{sec:isolated_pairs} and \ref{sec:galaxy_properties}.

\subsection{Results}\label{sec:results_all_pairs}
\begin{figure}[ht]
\centering
\includegraphics[width=0.5\textwidth]{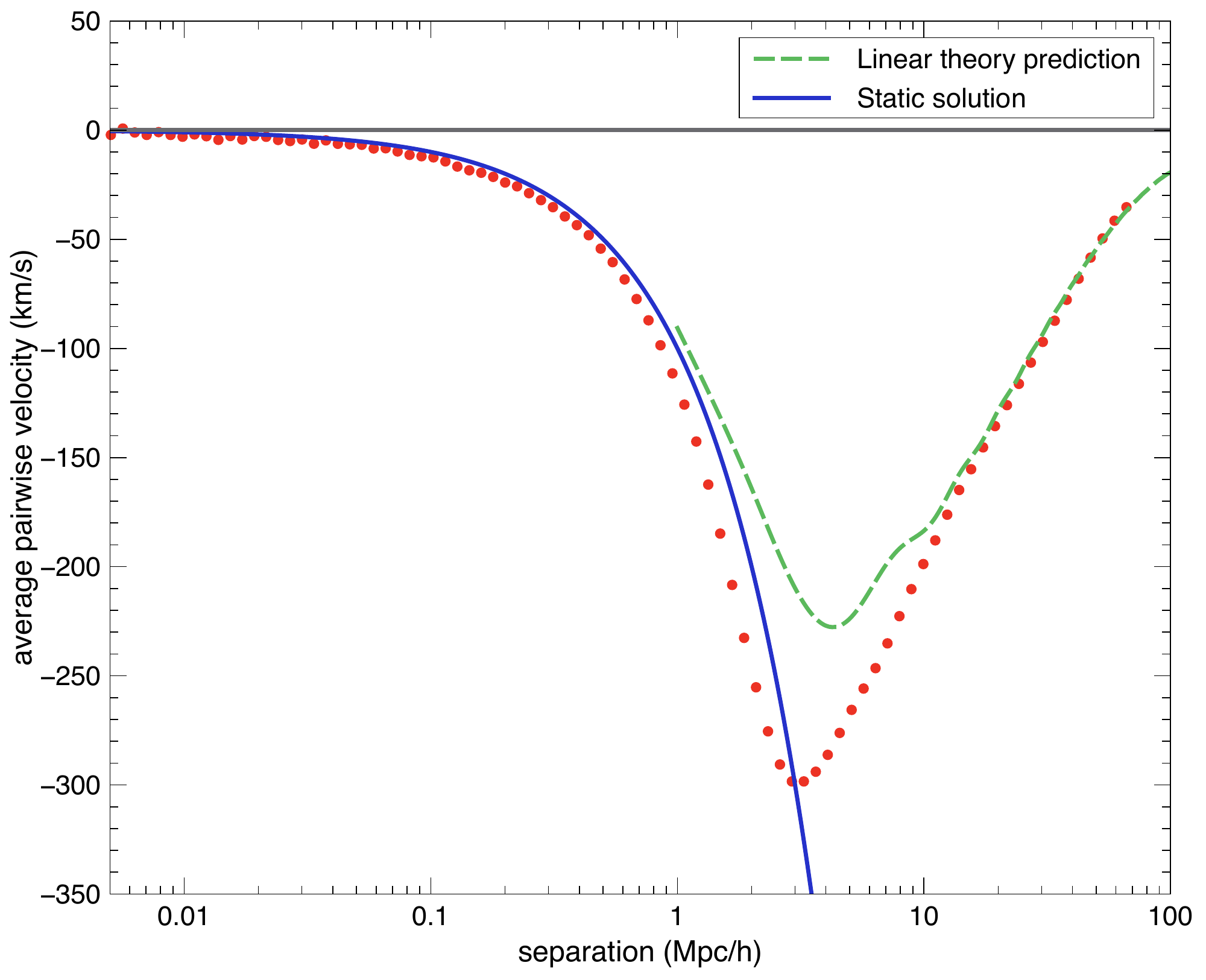}
\caption{Average relative velocity for all galaxy pairs at redshift $z=0$. The blue solid line represents the static solution, followed by pairs that have already virialized and do not feel the background Hubble flow. The green dashed line represents the linear theory model for pairwise velocities according to the prescription found in References \cite{fisher:1995a, reid:2011a} with a bias of $b=1$. Error bars, which are too small to be clearly seen are plotted at the one sigma confidence limit, assuming Poisson statistics.}
\label{fig:generic_all_pairs}
\end{figure}

In \fref{fig:generic_all_pairs} we show the average pairwise velocity $ v_{12} $ of all the galaxies within the \citet{guo:2011a} semi-analytic model at redshift $ z=0 $, as a function of separation $ d $.  The velocity curve is represented by the red dots with one-sigma error bars, while the blue and black lines are, respectively, the static and comoving solutions. %

We also plot the prediction of linear perturbation theory for $ v_{12} $ as the green dashed line, obtained using the prescription from \cite{fisher:1995a, reid:2011a}:
\begin{equation}
  v_{12}(d) = -\frac{fb}{\pi^2}\int dk\ k\ P_m(k)j_1(kd),
\end{equation}
where we have chosen a bias of $b=1$, which is in agreement with that measured from clustering within the galaxy catalogue.  

For separations larger than  $d>\mpch{10}$, we can see that the average peculiar velocity is correctly predicted by linear theory.  As we follow the velocity curve into the nonlinear regime, galaxy pairs approach the static line until they cross it at $d\sim\mpch{3}$.  This crossing marks the beginning of the infall regime, where the galaxies in the pairs get closer to each other, but with smaller velocities as their separation decreases.  On the smallest scales, the pairs asymptote to the static solution.

To use galaxy pairs as tracers of the cosmological expansion, we need their peculiar velocity to be small with respect to the Hubble flow or modellable.  A smaller correction is required if $ v_{12} $ is closer to zero comoving velocity than to the static solution. As we noted above, only for $d>\mpch{10}$ are the velocities closer to the comoving solution than the static solution, which is the regime of linear perturbation theory.  On scales $d\lesssim\mpch{3}$, galaxy pairs follow closely the static solution on average.  Their peculiar velocity component is equal and opposite to the Hubble flow, therefore they do not carry any cosmological information (refer to Appendix \ref{app:bound_systems}).

\section{Isolated pairs} \label{sec:isolated_pairs}
In this section we investigate how the dynamics of galaxy pairs changes when an isolation criterion is imposed and how this depends on the isolation radius, the allowed number of galaxies within this radius, and redshift. We use the same galaxy sample as in \sref{sec:all_pairs}.

\subsection{Method}\label{sec:method_isolated_pairs}
We initially define a galaxy pair to be isolated within a radius \riso if each galaxy in the pair has exactly 1 neighbor within \riso, and that neighbor is the other galaxy in the pair.  This is equivalent to drawing two spheres of radius \riso centered on the galaxies, and imposing the absence of galaxies extraneous to the pair in each of the spheres.  We shall weaken this requirement, allowing for the maximum number of neighbors \Nn in each sphere to be larger than 1.  Thus we can interpolate between the dynamics of galaxy pairs in the fully isolated case ($ \Nn = 1 $) and in the unconstrained case ($ \Nn \rightarrow \infty $). We implement such isolation criterion with arbitrary number of neighbors in a two-step process.  We first determine the number of neighbors for each galaxy in the simulation, and later use this information to select only those pairs where each galaxy has less than \Nn neighbors.

We fix the isolation radius to $ \riso = \mpch{4} $, which matches the definition of isolation adopted by \citet{marinoni:2010a}.  Note that in the plots that follow, we only look at separations less than the isolation radius to ensure that the pair is truly isolated.

\subsection{Results}
\begin{figure}[ht]
\centering
\includegraphics[width=0.5\textwidth]{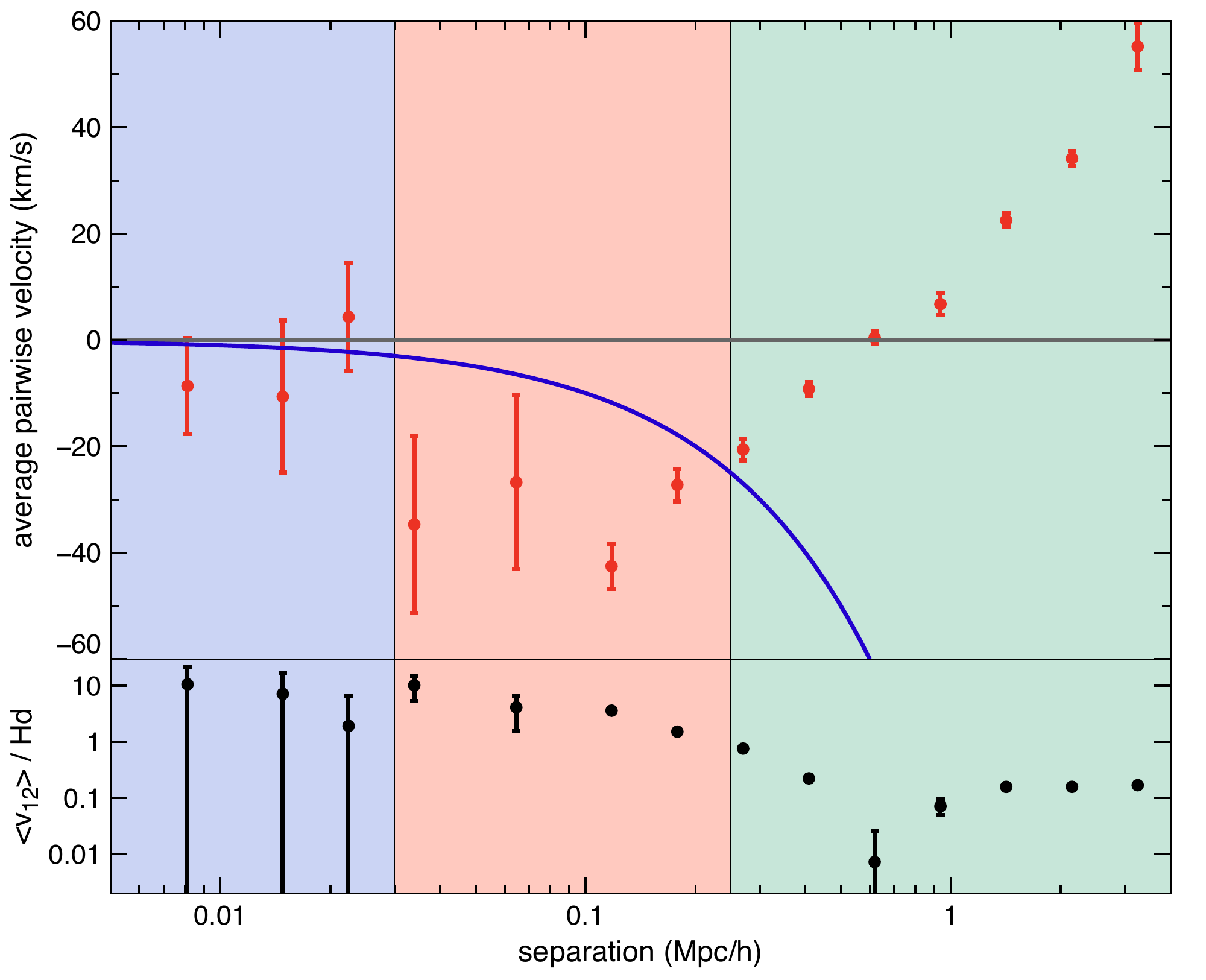}
\caption{Upper panel: average pairwise velocity for isolated galaxy pairs at redshift $z=0$. The isolation radius is taken to be \mpch{4} for each member of the pair. The blue solid line represents the static solution, showing the virialization of pairs. The three shaded regions represent three different regimes. The left blue area represents the virialization regime: galaxies within these separations have virialized and do not experience the background expansion. The middle red area shows the infall regime, where galaxies start to collapse to form bound systems. The right green region shows what we denote as the `void effect': on average, the isolated pair feels a stronger gravitational pull separating the pair rather than making it closer. The error bars shown are the one sigma confidence limit, assuming Poisson statistics. Lower panel: ratio of average pairwise velocities to the static solution $Hd$. Note that the y axis of this panel is in logarithmic scale. The error bars shown are the propagated one-sigma errors from the upper panel.}
\label{fig:generic_isolated}
\end{figure}

In the upper panel of \fref{fig:generic_isolated} we present the relative motion of galaxy pairs as a function of their separation for galaxies that are isolated within a \mpch{4} radius. The galaxies are taken from the \citet{guo:2011a} semi-analytic model applied to the Millennium simulation at redshift $ z = 0 $. The data points are plotted in red with one-sigma error bars. The blue line is the static solution and the black line the comoving solution -- see \sref{sec:method_allpairs} for their definition.

Our first comment on \fref{fig:generic_isolated} regards the error bars, which are much larger than those in the non-isolated case of \fref{fig:generic_all_pairs}.  The reason is that the imposition of an isolation criterion results in a drastic reduction of the galaxy pairs found, which in the case of \fref{fig:generic_isolated} are only $ 694 $
\footnote{Equivalent to one isolated pair every $10^6$ other pairs for $ d = \mpch{1} $}.
This number is in line with \citet{marinoni:2010a}, who find $ 721 $ pairs for their low-redshift SDSS sample, and $ 509 $ for their DEEP2 sample. 

The most striking feature about the dynamics of isolated pairs is the roughly logarithmic growth of the peculiar velocity $ v_{12} $ for scales larger than $ \sim \mpch{0.2} $.
The behavior of $ v_{12} $ can be explained when we recognize that the dynamics is determined by the combined effect of two competing forces: the mutual attraction between the two galaxies, dominant for small separations, and the disrupting outflow from the void -- the \emph{void effect}, dominant for separations close to the isolation radius.
For small separations, the mutual attraction of the members of the pair overcomes the void effect and we see an infall regime.  As we study objects with larger separations, the void effect becomes dominant and we see a logarithmically growing pairwise velocity $ v_{12} $.

In the lower panel of \fref{fig:generic_isolated} we plot $ v_{12}/H d $, that is the ratio of peculiar velocity to Hubble flow. For separations $ 0.4 < d < \mpch{4} $ we find an almost comoving regime where the peculiar velocities are less than $ 20\% $ of the Hubble flow.  In such a regime,  the RSD corrections are small, and one could hope that they are easier to model so that $ v_{12} $ becomes a proxy of the expansion rate.  In the following subsections we shall investigate how the comoving regime depends on the isolation radius, the local density and redshift.

\subsubsection{Varying the isolation radius}\label{sec:shrinking}
\begin{figure}[ht]
\centering
\includegraphics[width=0.5\textwidth]{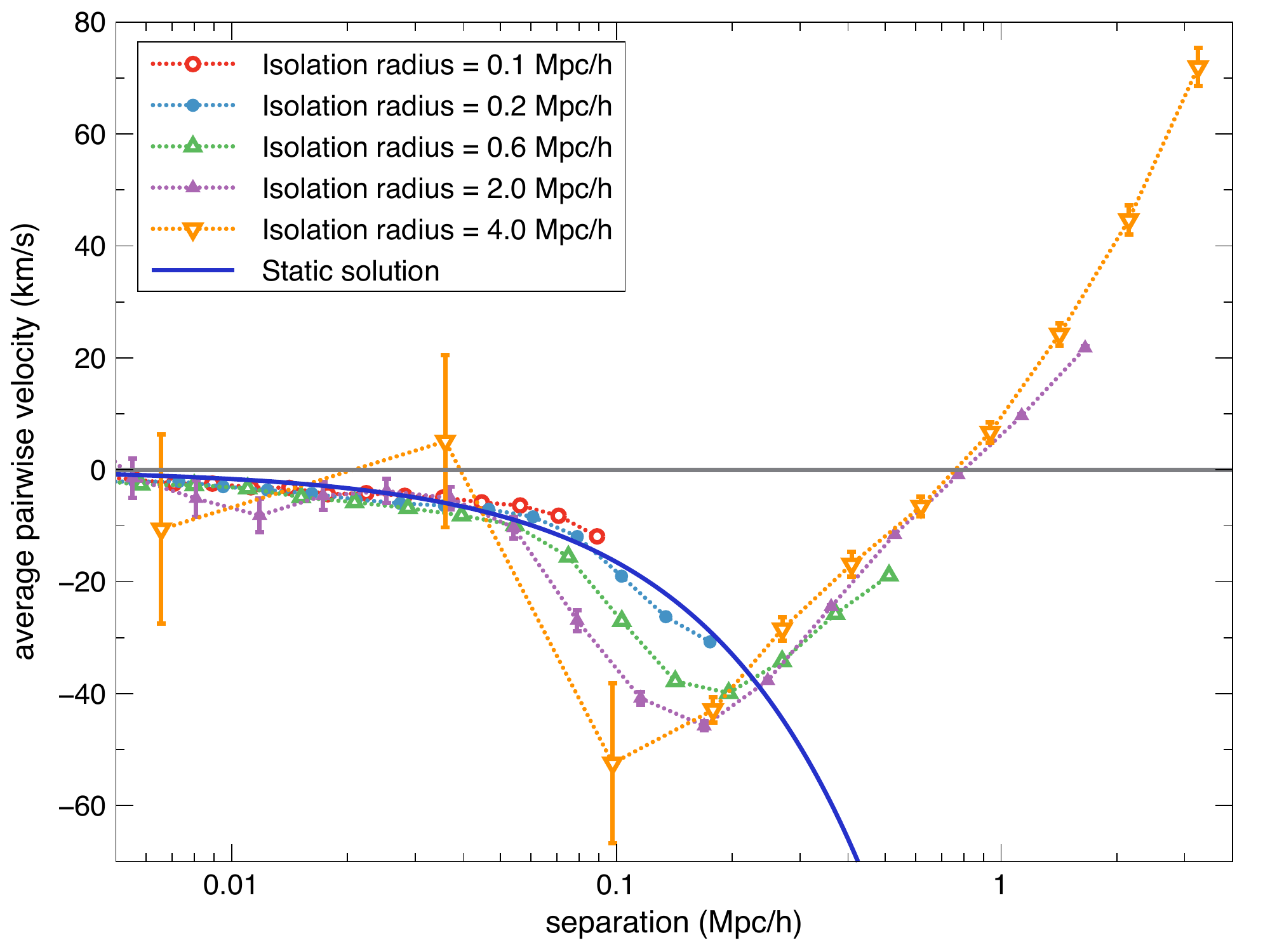}
\caption{Average pairwise velocity for isolated galaxy pairs at redshift $z=0.989$ for different isolation radii ranging from \mpch{0.1} to \mpch{4}.}
\label{fig:isolation_radii}
\end{figure}

It is interesting to investigate whether the void effect, giving the approximately logarithmic growth of $ v_{12}(d) $ for isolated pairs, is still present when we allow the size of the isolation radius to vary.  We demonstrate this in \fref{fig:isolation_radii}, where we show the peculiar velocities of galaxy pairs with $ \riso $ varying from \mpch{0.1} to \mpch{4}, at redshift $ z = 0.989 $.

For $ \riso \geqslant \mpch{0.6} $ the presence of the void severely affects the dynamics of galaxy pairs.  The logarithmic growth of the peculiar velocity is visible, even though it is just a hint for the $ \riso = \mpch{0.6} $ data points.  The cosmological scale $ d_0 $, defined as the separation where the peculiar velocity $ v_{12} $ vanishes, occurs at $ d_0 \simeq \mpch{0.8} $ and appears to be independent of the isolation radius.  

At small separations, we cannot see any noticeable differences between the various $ \riso $ datasets.  They all follow the static solution line for $d<\mpch{0.05}$, suggesting that isolated galaxy pairs tend to virialize on the smallest scales just as non-isolated ones do.

The number of isolated galaxy pairs at $ z=0.989 $ increases from $ 435 $ to $ 71,201 $ when reducing the isolation radius from $ 4 $ to \mpch{2}, a factor of roughly $ 160 $.  As we have noted above, the $ \riso = \mpch{2} $ pairs still experience a regime where peculiar velocities are negligible with respect to the Hubble flow.  Hence their velocity difference still traces the cosmological expansion, although the maximum separation for which this is the case is halved with respect to the $ \riso = \mpch{4} $ case.  We conclude that using pairs isolated within a \mpch{2} radius as cosmological tracers would drastically reduce the statistical error with respect to the \mpch{4} case, while still needing minimal corrections for RSD, provided that the cosmological dependence of the RSD could be modeled. We shall discuss this in more detail later.

\subsubsection{Varying the isolation density criterion}\label{sec:populating}
Here, we investigate how the dynamics of galaxy pairs changes if we relax the isolation criterion by increasing the allowed number of galaxies $ \Nn $ within the isolation sphere of \mpch{4}.

\begin{figure}[ht]
\centering
\includegraphics[width=0.5\textwidth]{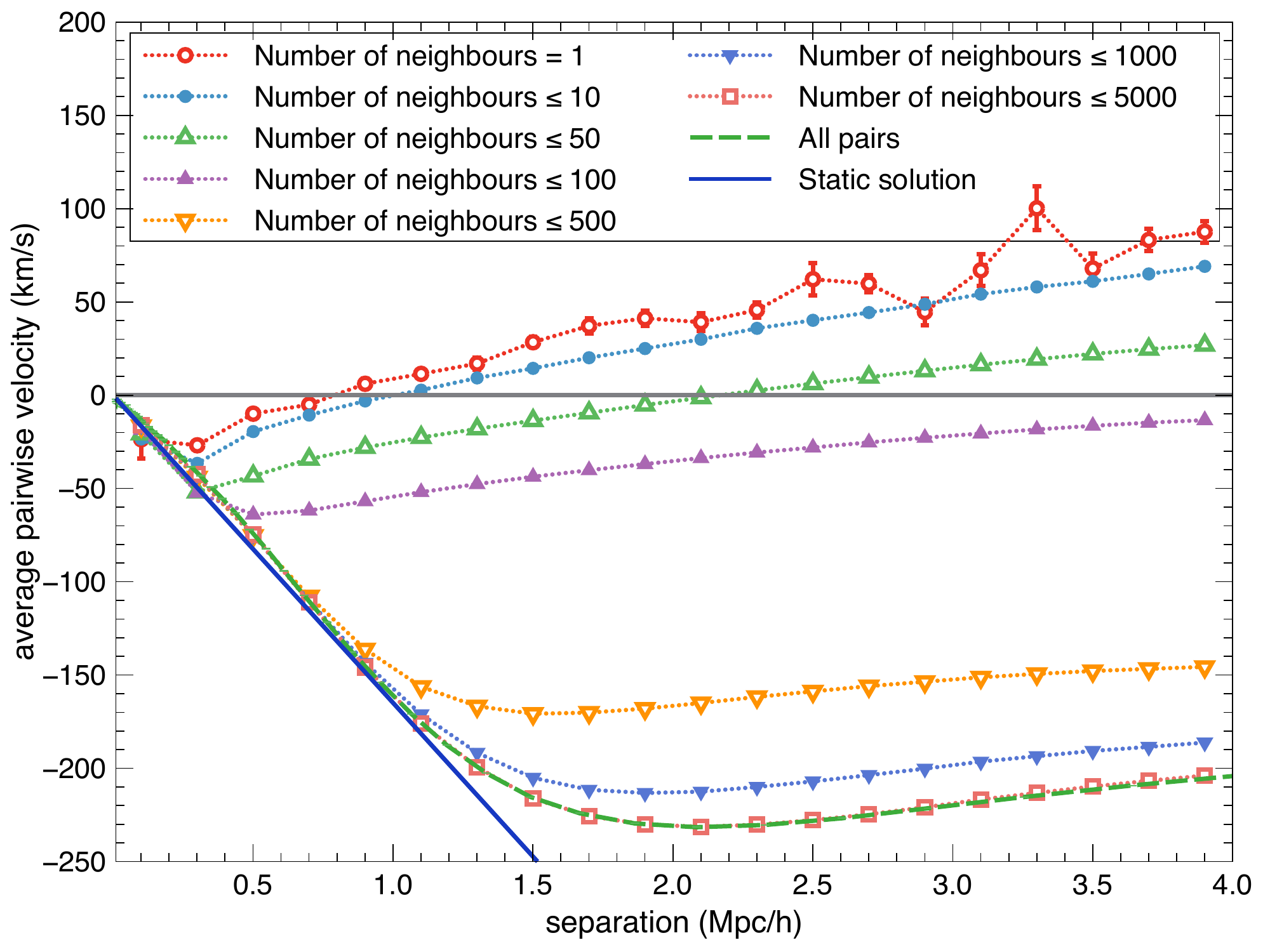}
\caption{Average pairwise velocity for isolated galaxy pairs at redshift $z=0.989$ for different number of neighbors within the isolation sphere of \mpch{4}. The blue solid line represents the static solution, showing the virialization of pairs. The green dashed line represents the average pairwise velocity for non-isolated pairs. The error bars shown are the one sigma confidence limit, assuming Poisson statistics.}
\label{fig:neighbours}
\end{figure}

In the linear plot of \fref{fig:neighbours}, we present the average peculiar velocity $ v_{12} $ at $ z=0.989 $ for 7 values of $ \Nn $ ranging from $ \Nn = 1 $ (equivalent to the pure isolated case of \fref{fig:generic_isolated}) to $ \Nn = 5000 $.  We also plot $ v_{12} $ for the non-isolated galaxy pairs, as already shown in \fref{fig:generic_all_pairs}, as a dashed green curve.  As $ \Nn $ increases, the different $ v_{12} $ curves monotonically fill the gap between the pure isolated case and the non-isolated case.  A good agreement between the dynamics of pairs with and without isolation criterion is reached once we allow each galaxy in the pair to have $ 5000 $ other neighbors.

In the $ \Nn = 10 $ case, we found $ 250,670 $ pairs, roughly a factor $ 600 $ more pairs than in the fully isolated case of $ \Nn = 1 $.  Nonetheless, the $ v_{12} $ curve for $ \Nn = 10 $ is strikingly similar to the $ \Nn = 1 $ one.  In particular, the void effect still seems to trigger the logarithmic growth of the peculiar velocity, with $ v_{12} $ crossing the zero line at $ d_0 \simeq \mpch{1} $ (in the fully isolated case, we have $ d_0 \simeq \mpch{0.8}) $. Hence, we suggest that pairs that are not completely isolated trace the cosmological expansion almost as well as the fully isolated ones, with the added benefit of a much better statistics.

\subsubsection{Varying redshift} \label{sec:varying_redshift}

\begin{figure}[ht]
\includegraphics[width=0.5\textwidth]{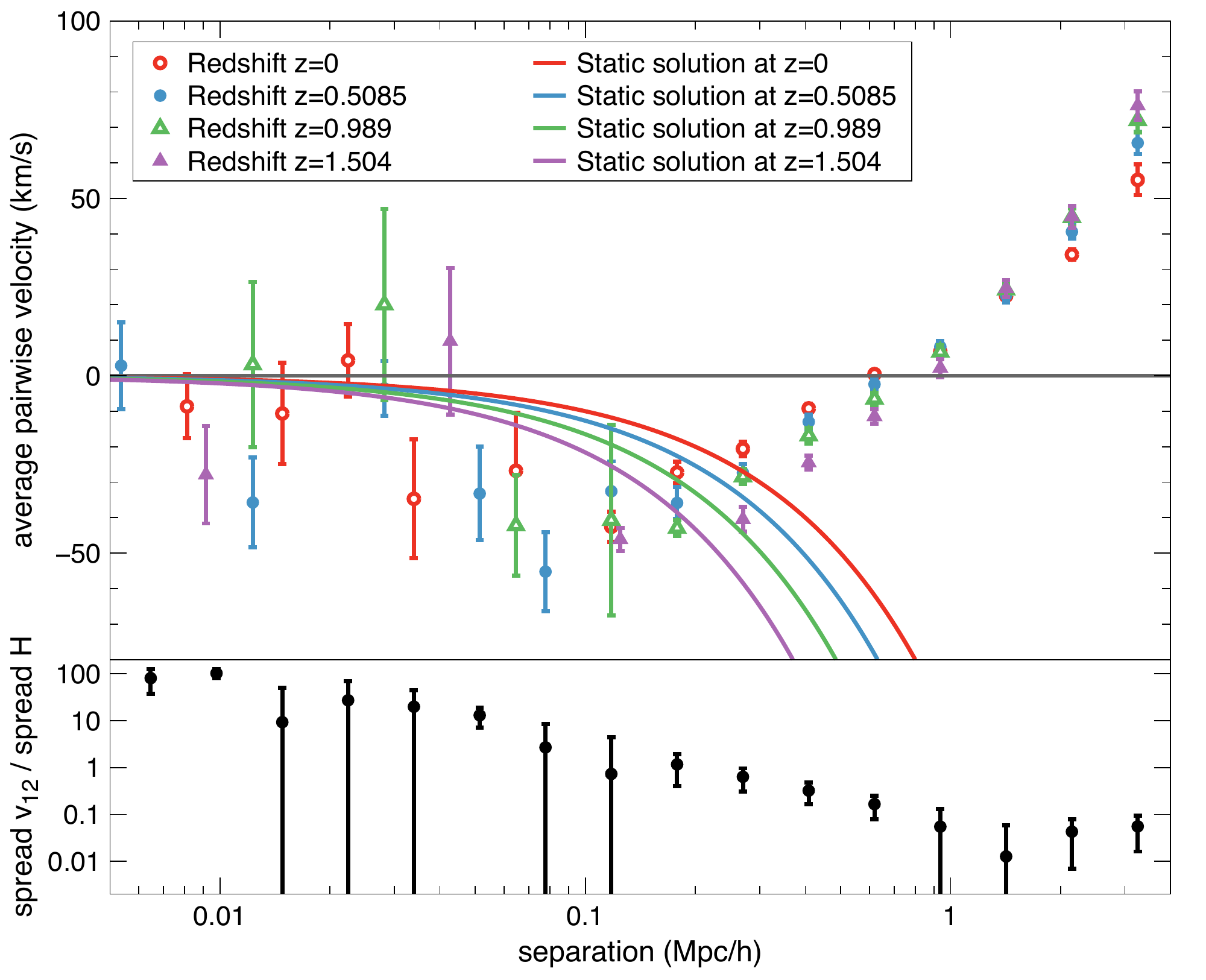}
\caption{Upper panel: variation of the average pairwise velocity of isolated galaxy pairs with redshift as a function of separation. The isolation radius is taken to be \mpch{4} for each member of the pair. Lower panel: range in pairwise velocity for each separation bin over range in the static solutions at each separation bin. Note that the y axis of this panel is in logarithmic scale. The error bars shown are the propagated one-sigma errors from the upper panel.}
\label{fig:iso_redshift_variation}
\end{figure}

We illustrate the redshift dependence of the peculiar velocity for isolated pairs in the top panel of \fref{fig:iso_redshift_variation}, where we plot $ v_{12}(d) $ for the four redshifts $ z = 0, 0.5085, 0.989, 1.504 $.  It is remarkable that for separations $ d \gtrsim \mpch{0.2} $ the peculiar velocity depends only slightly on redshift.  The scale $ d_0 $, defined as the separation where $ v_{12} $ vanishes, ranges from \mpch{0.6} at $ z=0 $ to \mpch{0.9} at $ z=1.504 $.  This is a small variation if we consider that at $ z=1.504 $ the Universe was at one third of its current age.

In the bottom panel of \fref{fig:iso_redshift_variation} we plot the ratio between the range in $ v_{12} $ and the range in $ H d $ at a given separation.  For separations $ 1 < d < \mpch{4} $, the change of peculiar velocity with redshift is only $ 10 \% $ of the change in Hubble flow.  This implies a weak dependence of $ v_{12} $ on cosmological expansion on those scales.

\subsection{Cosmological implications}\label{sec:cosmological_implications}
\fref{fig:generic_isolated} shows that for $ 0.4 < d < \mpch{4} $ isolated galaxy pairs at $ z=0 $ are nearly comoving with the Hubble flow.  Thus, they move with the cosmological expansion and, for this cosmology and epoch, only need a small RSD correction. Using different redshift slices as a way to test different cosmological expansion rates, in the lower panel of \fref{fig:iso_redshift_variation}, we identify a second regime for $ 1 < d < \mpch{4} $ where the peculiar velocity $ v_{12}(z) $ depends only slightly on the cosmological expansion. This suggests the intriguing possibility of isolated pairs behaving in the same way on those scales regardless of the assumed cosmology. In particular, the lower panel of \fref{fig:iso_redshift_variation} shows that the variation in RSD model is less than the variation in expansion rates. Thus we conclude that there is cosmological signal to be extracted here.

We refer to the intersection of these regimes, where we have almost comoving pairs with small redshift evolution, as a cosmological regime, since we might be able to use these pairs as cosmological tracers. Measuring galaxy pairs in the cosmological regime would still induce a systematic error due to the fact that peculiar velocities are non-zero.  As the correction is of the $10\%$ level -- see lower panel of \fref{fig:generic_isolated} -- and we might suppose to be able to model this at the same level, we would have a $1\%$ systematic correction to contend with.  This claim is little more than a speculation at this stage; in order to falsify or confirm it, one needs to model isolated pairs in detail and to analyze N-body simulations with different underlying cosmologies.

Pairs isolated within a radius of \mpch{4} are rare objects -- see \fref{fig:number_pairs} -- and this might result in significant statistical error when dealing with observations.  In Sections \ref{sec:shrinking} and \ref{sec:populating} we found that one can drastically increase the number of pairs while keeping the RSD correction small by either reducing the isolation radius to \mpch{2} or allowing up to $ 10 $ galaxies to be neighbors of the pair.

\section{Varying galaxy properties} \label{sec:galaxy_properties}
The main result of the previous section, illustrated in Fig.\ \ref{fig:generic_isolated} and \ref{fig:iso_redshift_variation}, is that there is a regime where isolated galaxy pairs may be used as tracers of expansion with correction for RSD that is weaker than the signal to be measured.  Such a finding relies on the ability of measuring the redshift of galaxies regardless of their mass or luminosity.  This is clearly not the case when dealing with actual galaxy surveys, whose flux sensitivity is limited.   To model such a selection bias, we need to investigate ways of selecting galaxies from simulations which mimic the selection process of surveys.   In this section we address this issue by forming subsamples where galaxies are selected according to subhalo mass, stellar mass and magnitude.  We also study the redshift dependence of our results by analyzing 4 different redshifts: $ z = 0,\,0.5085,\,0.989,\,1.504 $.

\subsection{Method} \label{sec:method_varying}
\begin{table*}[t]
    \begin{tabular}{ r*{8}r }
        $\np$  &\:\:\:  $m(\msun/h)$  &\:\:\: $\rmag$  &\:\:\:  $\smass(\msun/h)$  &\:\:\:  n(z=0) (Mpc$^{-3}$)  &\:\:  \%(z=0)  &\:\:\:  \%(z=0.5085)  &\:\:\:  \%(z=0.989)  &\:\:\:  \%(z=1.504)  \\
        \hline \hline
        20   & \sci{1.72}{10} &        &                & \sci{8.1}{-2} &   100  &   100  &   100  &   100 \\
        100  & \sci{8.6}{10}  & -19.27 & \sci{1.40}{9}  & \sci{1.4}{-2} & 17.10  & 17.81  & 17.77  & 17.49 \\
        500  & \sci{4.3}{11}  & -21.72 & \sci{1.59}{10} & \sci{3.1}{-3} &  3.85  &  3.73  &  3.45  &  3.09 \\
        1000 & \sci{8.6}{11}  & -22.17 & \sci{2.60}{10} & \sci{1.6}{-3} &  2.01  &  1.88  &  1.67  &  1.41 \\
        5000 & \sci{4.3}{12}  & -22.86 & \sci{5.36}{10} & \sci{3.4}{-4} &  0.42  &  0.35  &  0.27  &  0.19
    \end{tabular}
    \caption{Cuts imposed on our galaxy sample from the \citet{guo:2011a} semianalytic model. The first line corresponds to no cut at all; in that case, for the particle number column, we report the \citet{guo:2011a} resolution limit of $ \np = 20 $.  The columns with a percentage sign denote the percentage of galaxies surviving the cut at a given redshift.  The \rmag and \smass columns refer to the cuts performed at $ z = 0.989 $, as it is the only redshift we plot for these quantities (see Figures \ref{fig:rmag_variation} and \ref{fig:stellar_mass_variation}).  Note that the \rmag cuts are intended to be upper limits, while the mass cuts are lower limits.}
\label{tab:cuts}
\end{table*}

\begin{figure*}[t]
\subfloat[No isolation]{\label{fig:npairs_noiso}\includegraphics[width=0.5\textwidth]{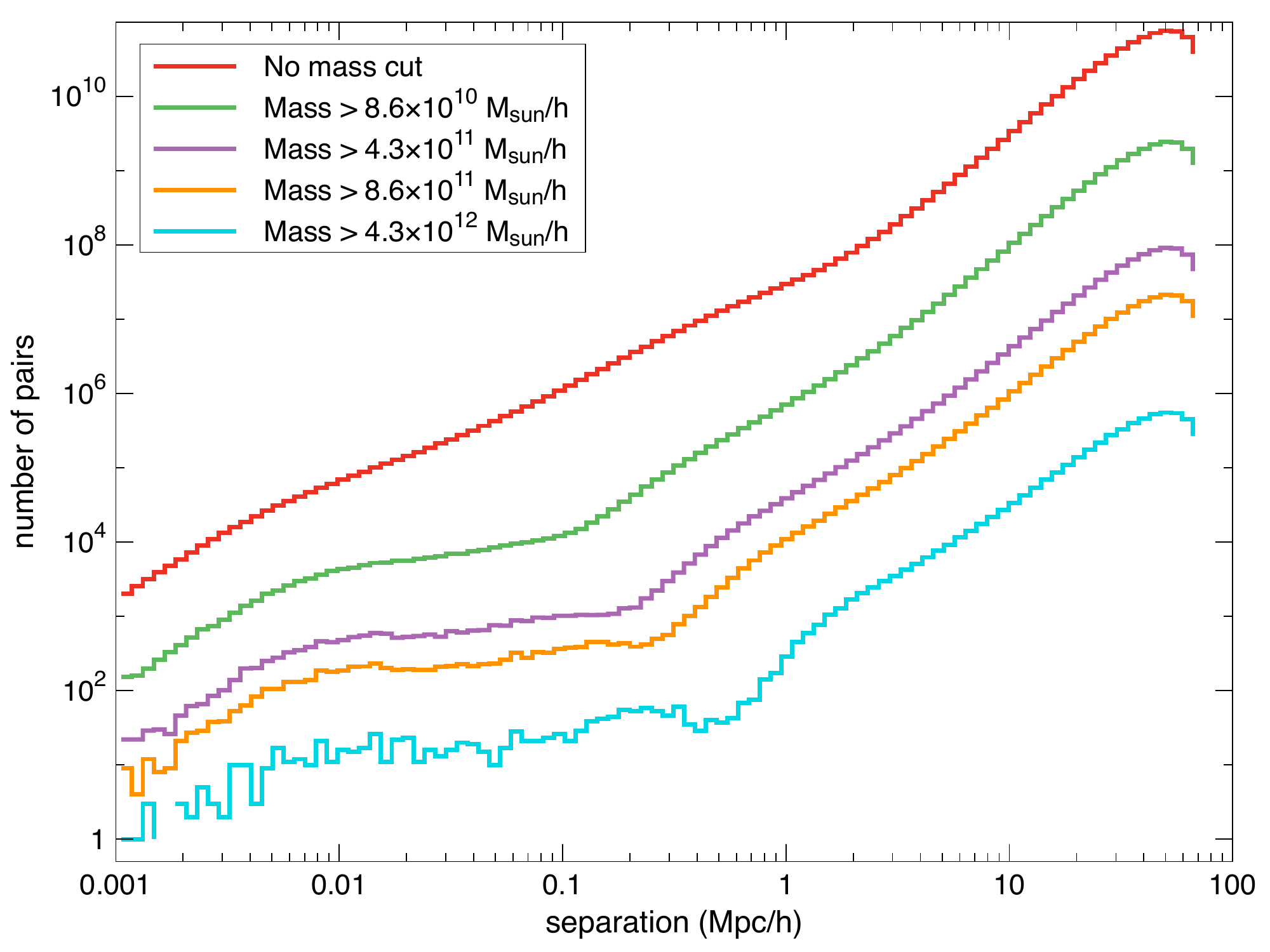}}
\subfloat[Isolation of \mpch{4}]{\label{fig:npairs_isolation4}\includegraphics[width=0.5\textwidth]{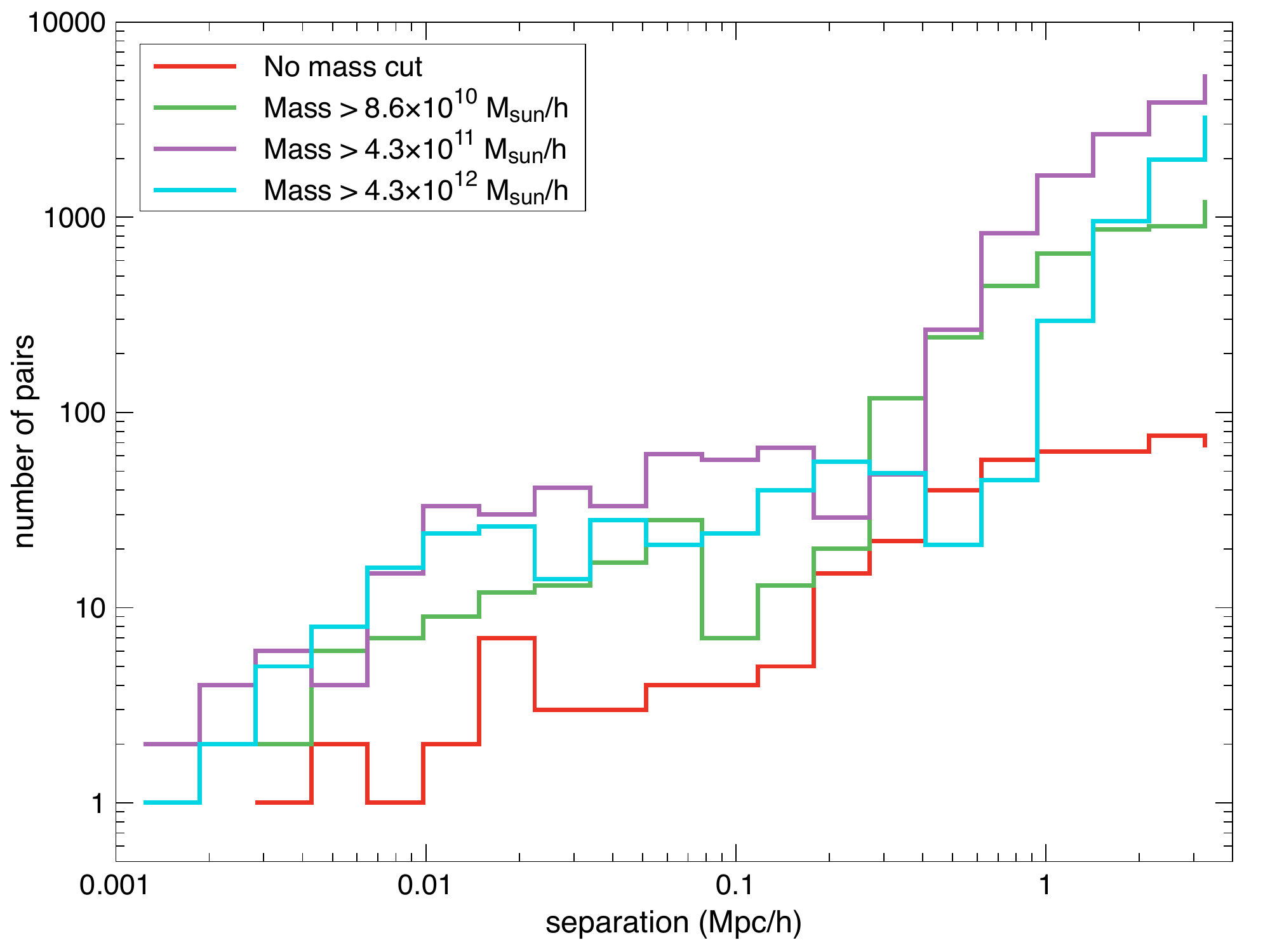}}
\caption{Number of galaxy pairs for different mass cuts as a function of separation at redshift $z=0.989$.  In the right panel we omitted the $ m > \msunh{8.6}{11} $ curve for the sake of readability.}
\label{fig:number_pairs}
\end{figure*}

We select galaxy subsamples from the semi-analytic models by applying cuts on galaxy properties.  We then study the dynamics of each subsample by applying the same analyses of Sections \ref{sec:all_pairs} and \ref{sec:isolated_pairs}.  Initially we look at the number \np of dark matter particles of the subhalo the galaxy is in%
\footnote{For reference, this is the \emph{np} field of the \emph{Guo2010a} database in the Millennium simulation servers.}.
We consider $ \np = 100,\,500,\,1000,\,5000 $ corresponding to masses $ m = \sci{8.6}{10},\,\sci{4.3}{11},\,\sci{8.6}{11},\,\msunh{4.3}{12} $.  Note that, where no cut is made, the number of particles in each subhalo is always $ \np > 20 $, corresponding to \msunh{1.72}{10}, since this is the threshold that defines a bound subhalo according to \citet{guo:2011a}. 

We choose the dark matter mass as our main cut because it is directly related to the pairwise velocity dynamics, which is the main subject of this paper.  In order to make a more direct link with observations, we also study the pairwise velocity statistics when varying the rest-frame $ r $-band magnitude%
\footnote{More precisely, this is the rest-frame total absolute magnitude in the SDSS r-band, corresponding to the \emph{r\_mag} field in the \emph{Guo2010a} database of the Garching mirror and to the \emph{r\_SDSS} field in the \emph{Font2008a} database of the Durham mirror.}
and stellar mass of galaxies.

The limits on $ r $-band magnitude (hereafter \rmag) and stellar mass (hereafter \smass) are chosen such that, for a given \np cut, the corresponding \rmag and \smass cuts yield the same number of surviving galaxies. \tref{tab:cuts} reports the values of the limits used, together with the resulting fraction of surviving galaxies at each redshift. We apply these cuts to the data sets before running the pair-finder algorithms.  Thus, a pair that is isolated within its subsample may not be isolated when considering the full catalog, \ie our isolation criterion is sample dependent.  This implementation is in line with an analysis of an actual galaxy survey, limited by these cuts.

In \fref{fig:number_pairs} we show how many galaxy pairs we find after imposing the cuts given in \tref{tab:cuts}.  The number of unconstrained pairs (left panel) decreases monotonically with increasing mass cut.  Note that the drop-off in the number of pairs at large separations is due to the size of the individual boxes we consider and has no physical meaning.  For isolated pairs (right panel), the number density increases as we increase the mass cut, as a higher mass cut results in a sparser distribution of galaxies where it is easier to find isolated pairs.  Only for our most stringent mass cut, that is for $ m>\msunh{4.3}{12} $, do we see a slight decrease in the density of isolated pairs due to the small number of high mass galaxies.

\begin{figure*}[ht]
\subfloat[Redshift z=0]{\label{fig:z_0}\includegraphics[width=0.5\textwidth]{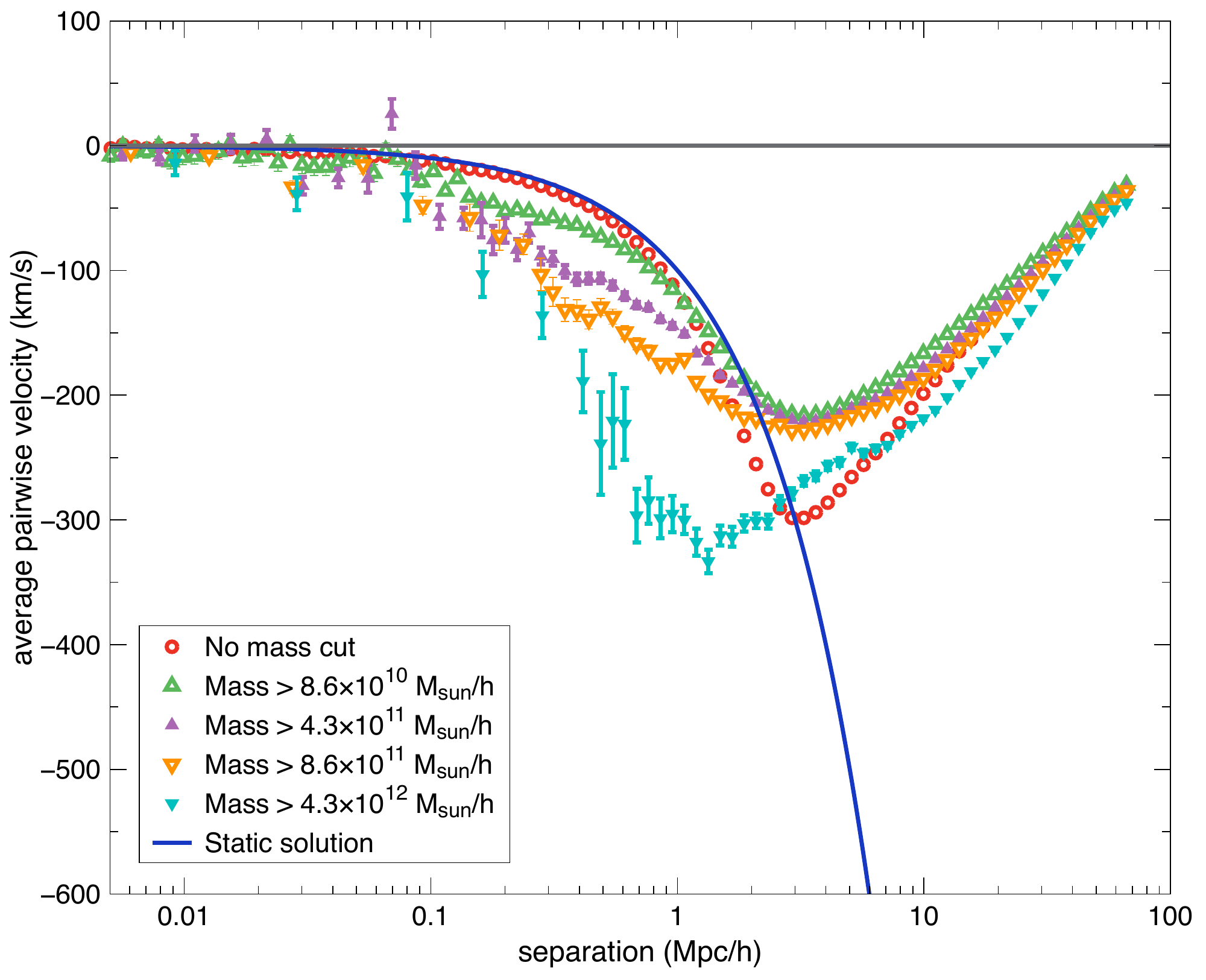}}
\subfloat[Redshift z=0.5085]{\label{fig:z_0.5}\includegraphics[width=0.5\textwidth]{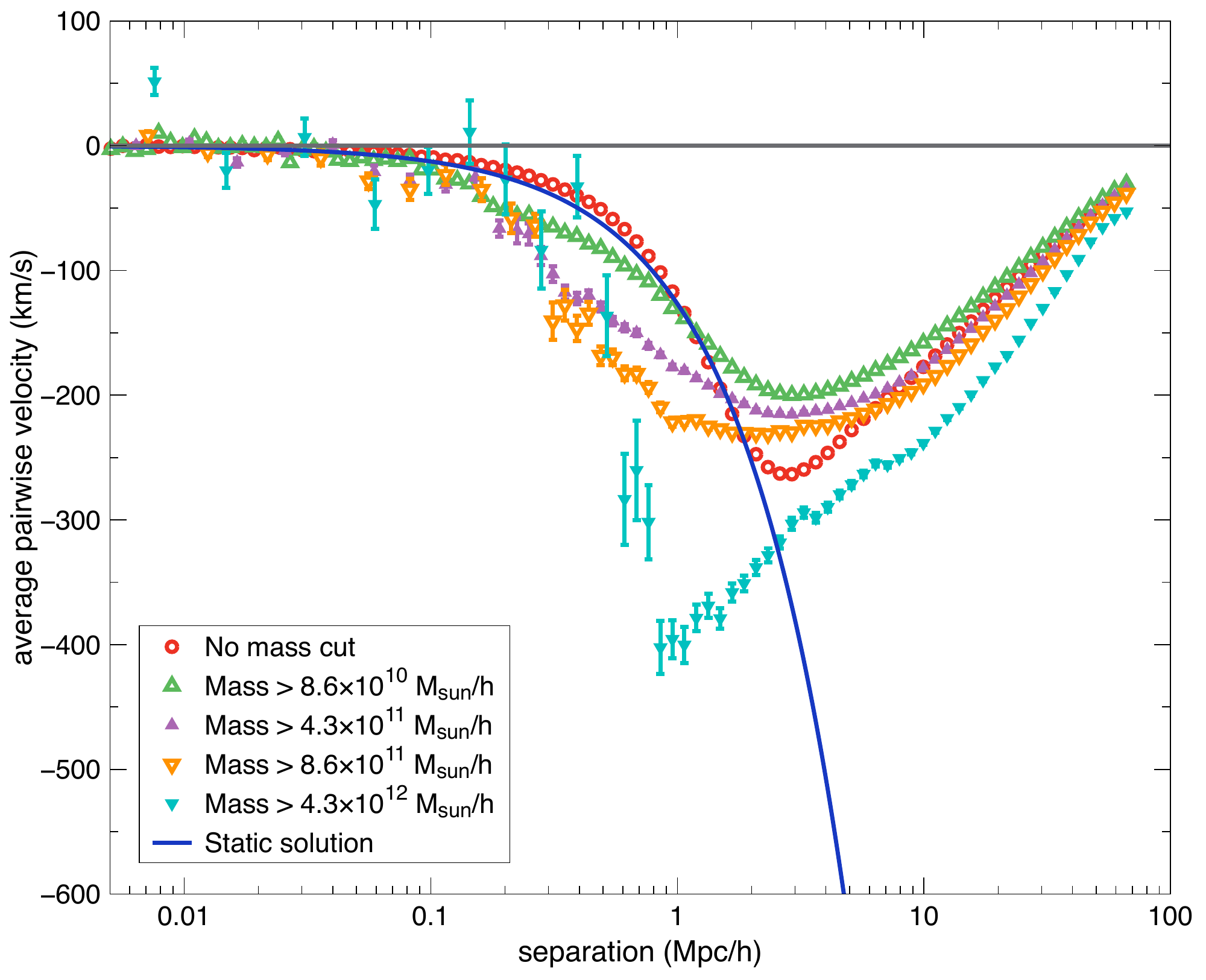}}\\
\subfloat[Redshift z=0.989]{\label{fig:z_1}\includegraphics[width=0.5\textwidth]{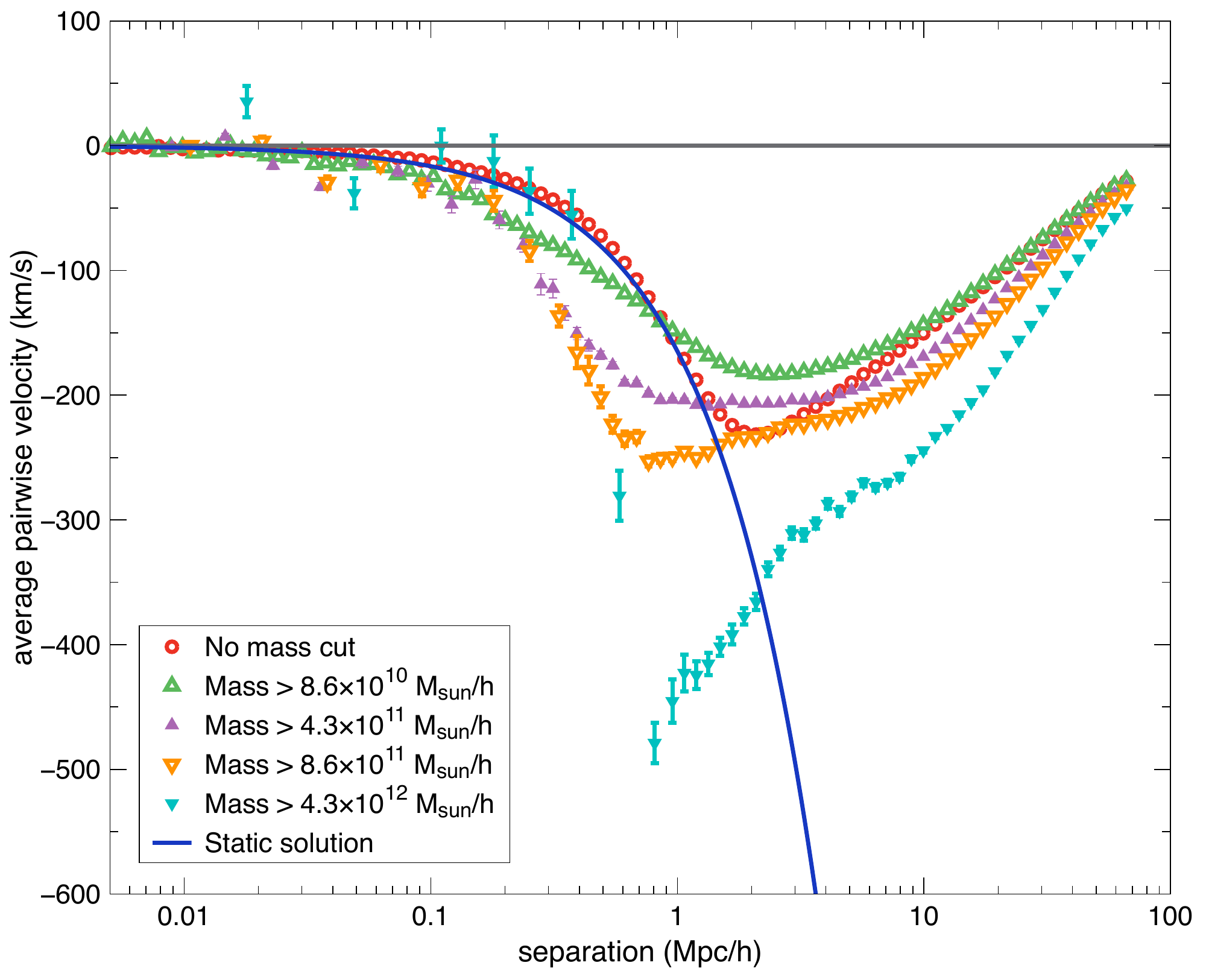}}
\subfloat[Redshift z=1.504]{\label{fig:z_1.5}\includegraphics[width=0.5\textwidth]{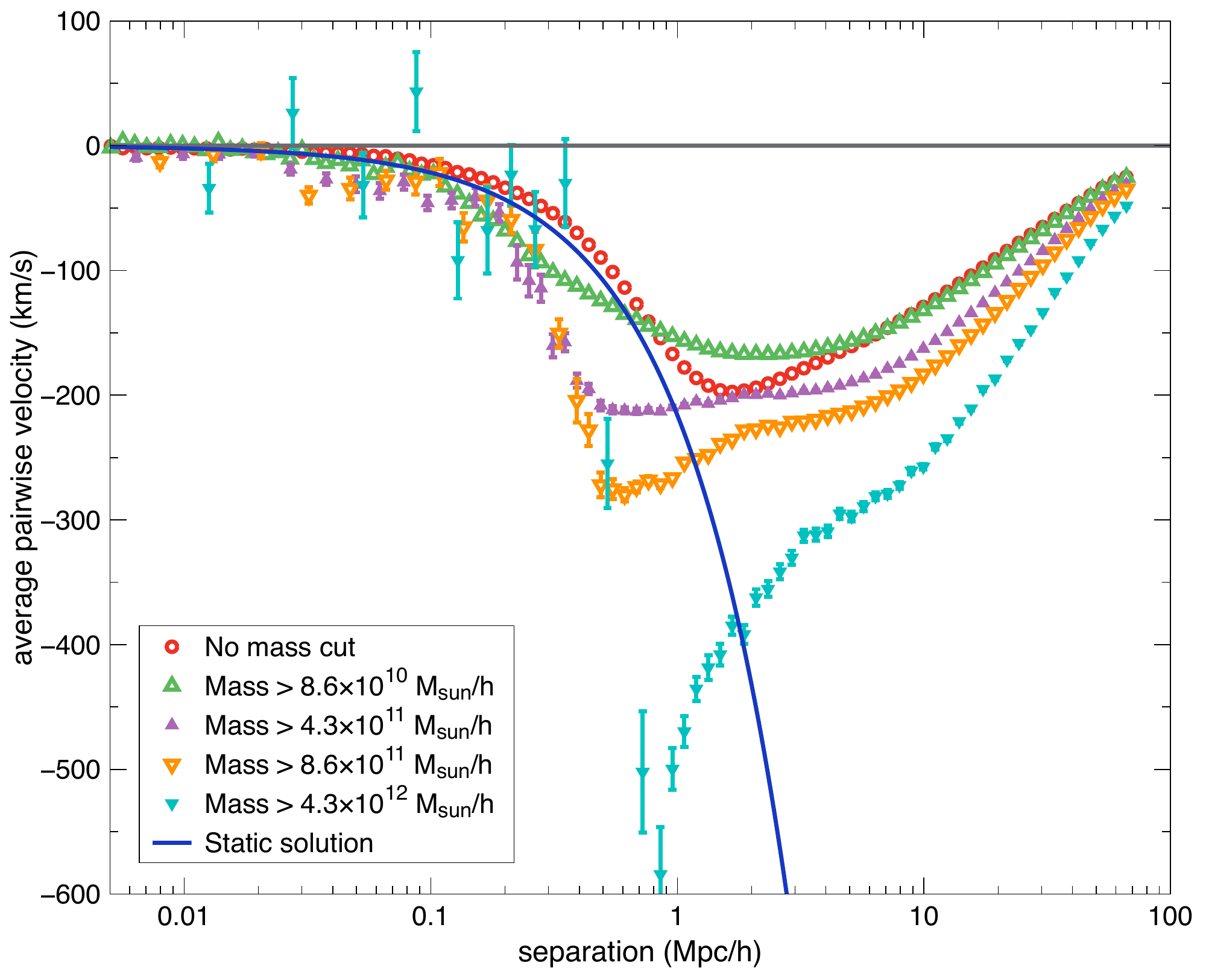}}
\caption{Variation of the average pairwise velocity of all galaxy pairs with redshift for different mass cuts (number of dark matter particles) as a function of separation.}
\label{fig:redshift_variation_noiso}
\end{figure*}

\subsection{Results}

\subsubsection{All galaxy pairs}
In \fref{fig:redshift_variation_noiso} we present the average pairwise velocity $ v_{12} $ for non-isolated pairs above different subhalo mass thresholds.  The four panels show the same selection procedure at different redshifts. The mass cuts range from \msunh{1.72}{10} up to \msunh{4.3}{12} as tabulated in \tref{tab:cuts}.  Note that the lowest mass cut corresponds to the smallest subhalo in the \citet{guo:2011a} semianalytic model, consisting of $ 20 $ dark matter particles.

The imposition of a mass cut has a significant impact on non-linear scales. Independent of redshift, \fref{fig:redshift_variation_noiso} shows that massive galaxy pairs experience an infall regime for separations smaller than $ d \lesssim \mpch{3} $.  While in such a regime, peculiar velocity increases with mass, with the most massive galaxies ranging from $ v_{12} \simeq \kms{330} $ at $ z=0 $ and $ v_{12} \simeq \kms{600} $ at $ z=1.504 $.  Lower mass galaxies, on the other hand, seem to follow the static solution up to higher separations, especially at low redshift. Our interpretation for such behavior is that galaxies in high mass pairs are more affected at small separations by their mutual attraction than the underlying density field.

\begin{figure*}[ht]
\subfloat[Redshift z=0]{\label{fig:z_0_iso}\includegraphics[width=0.5\textwidth]{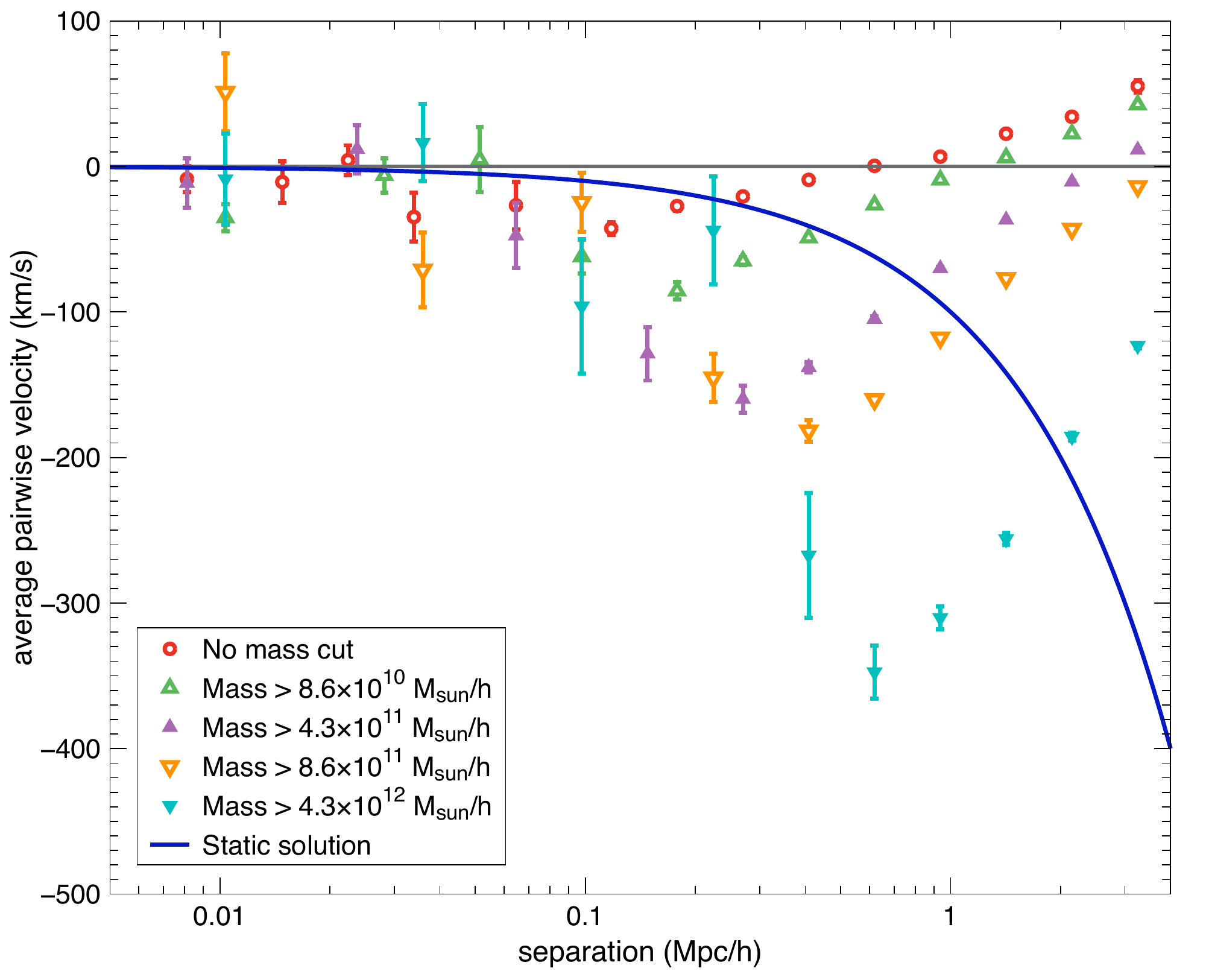}}
\subfloat[Redshift z=0.5085]{\label{fig:z_0.5_iso}\includegraphics[width=0.5\textwidth]{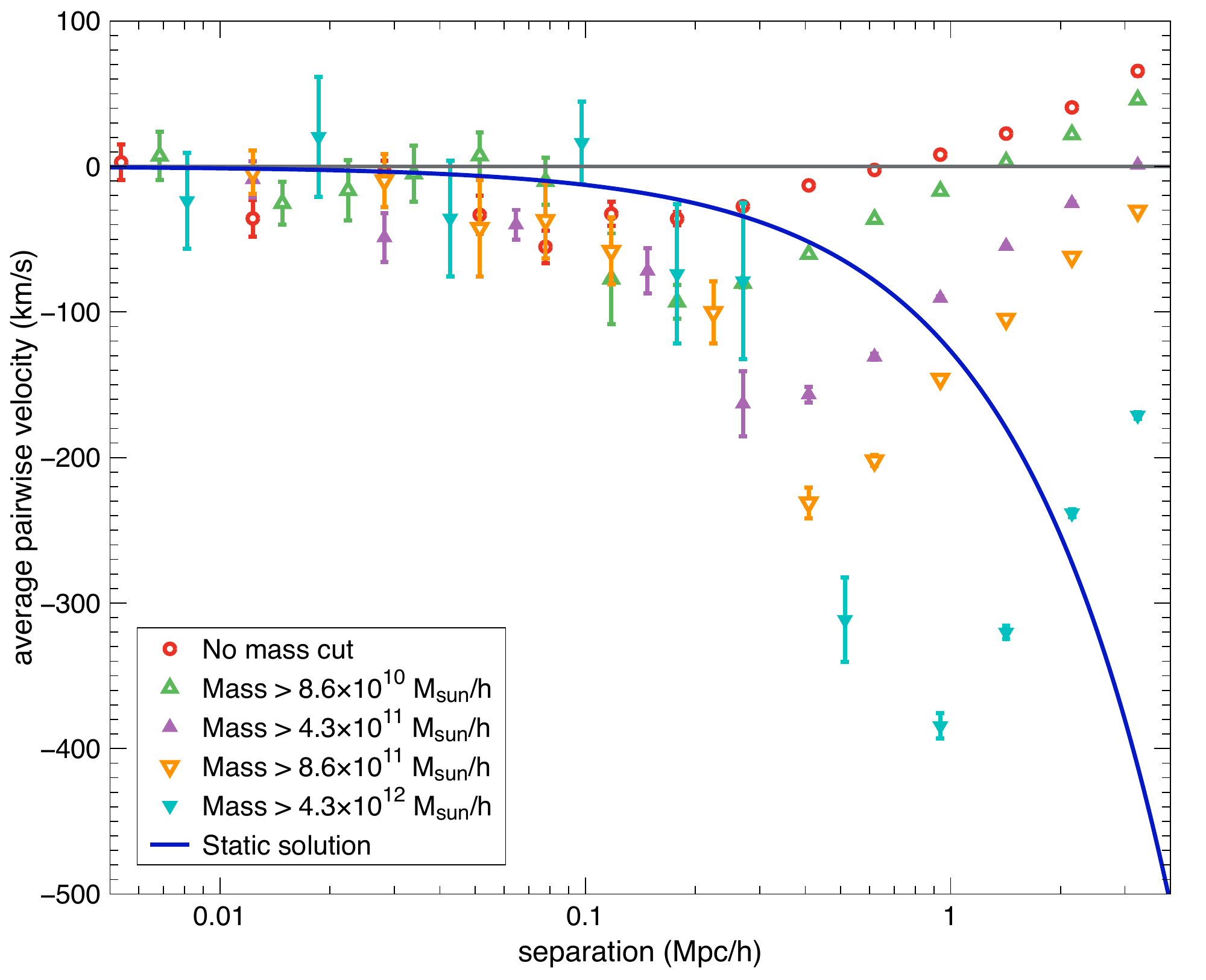}}\\
\subfloat[Redshift z=0.989]{\label{fig:z_1_iso}\includegraphics[width=0.5\textwidth]{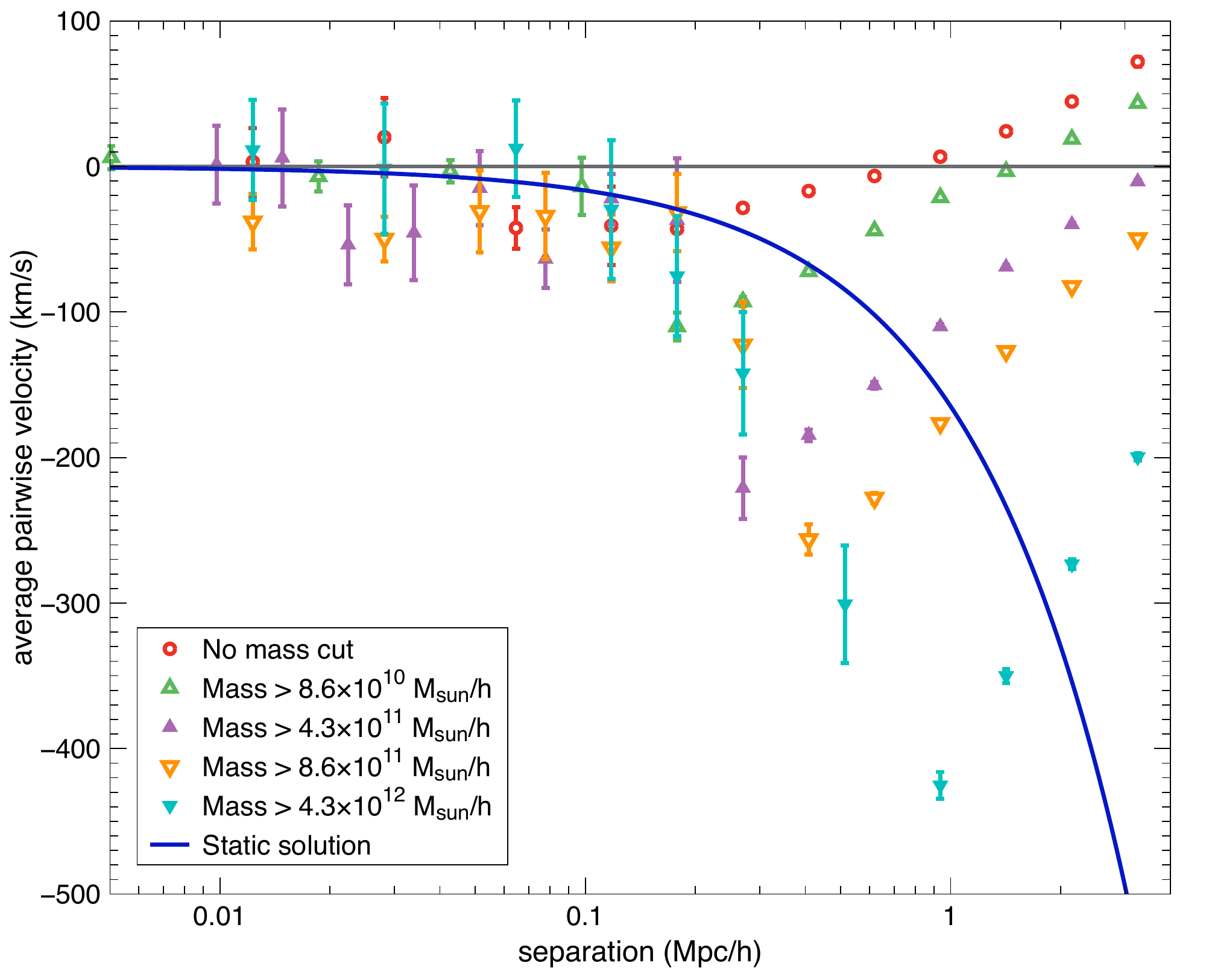}}
\subfloat[Redshift z=1.504]{\label{fig:z_1.5_iso}\includegraphics[width=0.5\textwidth]{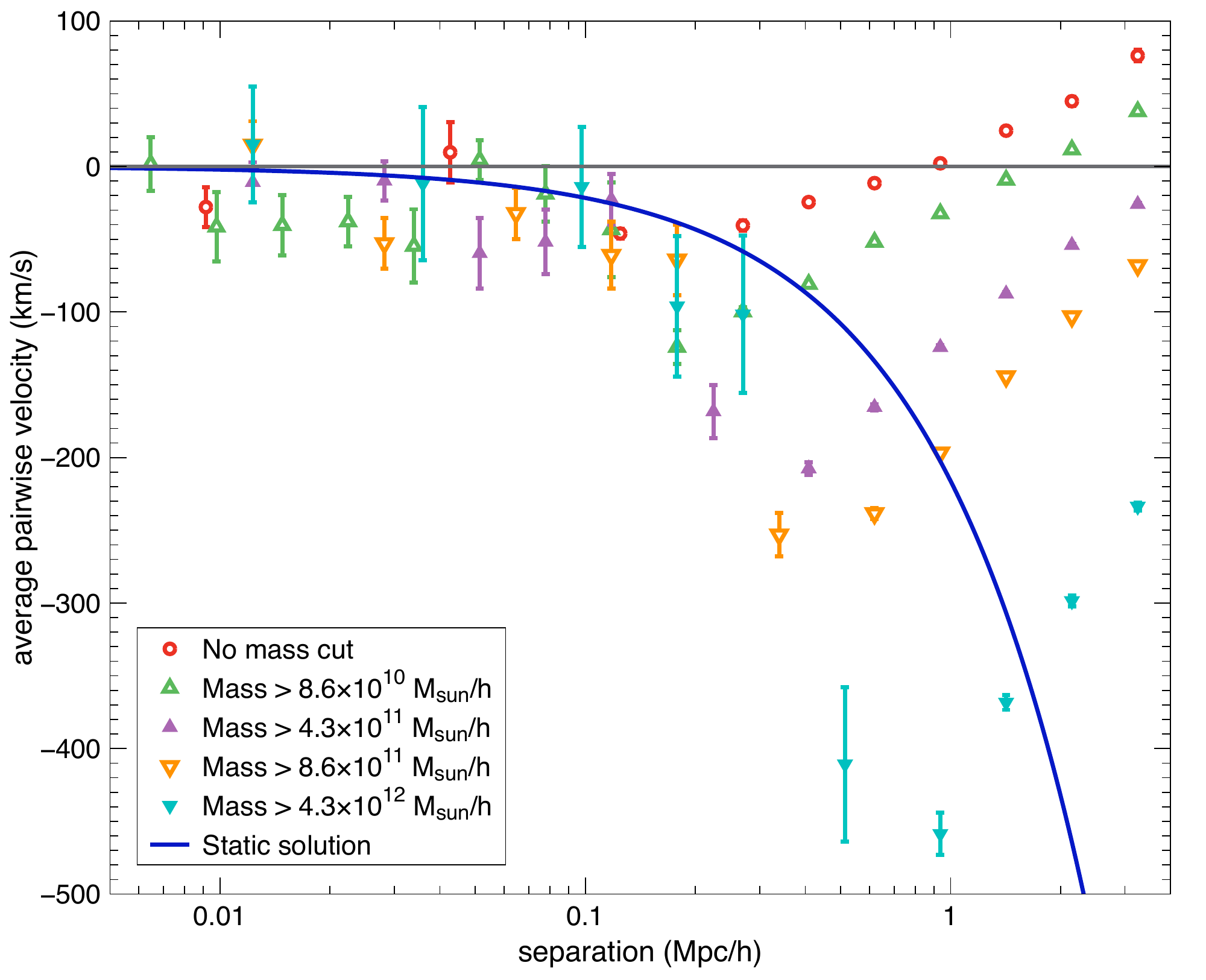}}
\caption{Variation of the average pairwise velocity of isolated galaxy pairs with redshift for different mass cuts (number of dark matter particles) as a function of separation. The isolation radius is taken to be \mpch{4} for each member of the pair.}
\label{fig:redshift_variation_isolation4}
\end{figure*}

\begin{figure}[h]
\includegraphics[width=0.5\textwidth]{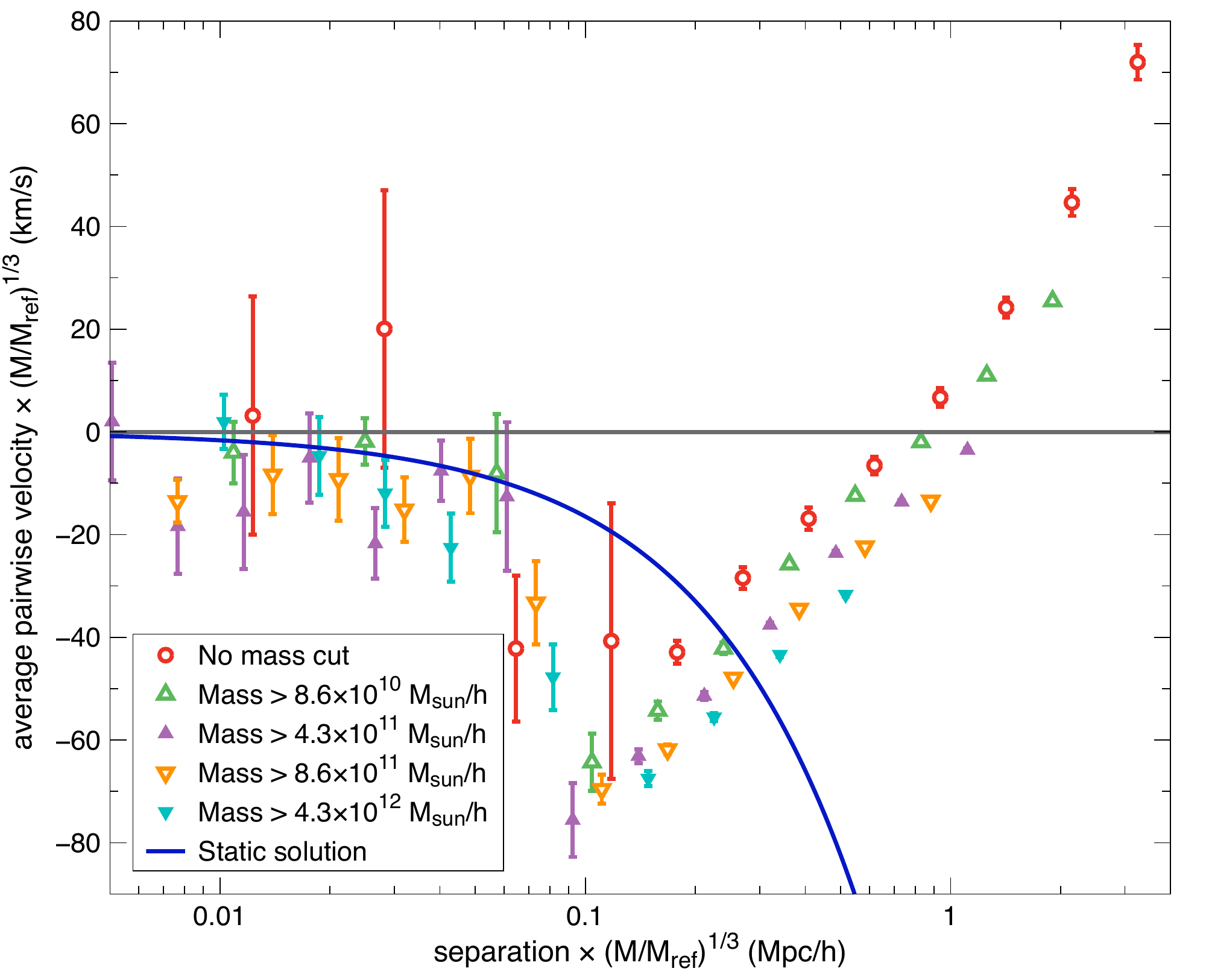}
\caption{Same plot as in \fref{fig:z_1_iso} where both the separation and the average pairwise velocity for each curve have been scaled by a factor $(M_{\text{ref}}/M)^{1/3}$ where $M_{\text{ref}}$ is the minimum mass of a subhalo: $M_{\text{ref}}=\sci{1.72}{10} \msun/h$. Note that the errors for each curve have been scaled accordingly.}
\label{fig:scaling_z1}
\end{figure}

For separations $ d > \mpch{10} $, the velocity curves at each redshift seem to converge to a common asymptote.  In \sref{sec:all_pairs}, we have shown that this limit is correctly predicted by linear theory -- see the agreement between the green curve and $ v_{12} $ in \fref{fig:generic_all_pairs}.  This means that, even though the non-linear dynamics of the different mass limit pairs differs, their behavior at large separations seems to be predicted by the same linear theory.

A remarkable feature of \fref{fig:redshift_variation_noiso} is the different redshift dependence of the various $ v_{12} $ curves.  The pairwise velocity of the heaviest galaxies (cyan triangles) greatly increases with redshift, while in the uncut case (red circles) it decreases.  The intermediate curves seem to experience smaller variations.

\subsubsection{Isolated pairs}

\begin{figure*}[ht]
\subfloat[\;No mass cut\;\;(\:$ \np \geqslant 20 \:)$]
 {\label{fig:np_0_iso}\includegraphics[width=0.5\textwidth]{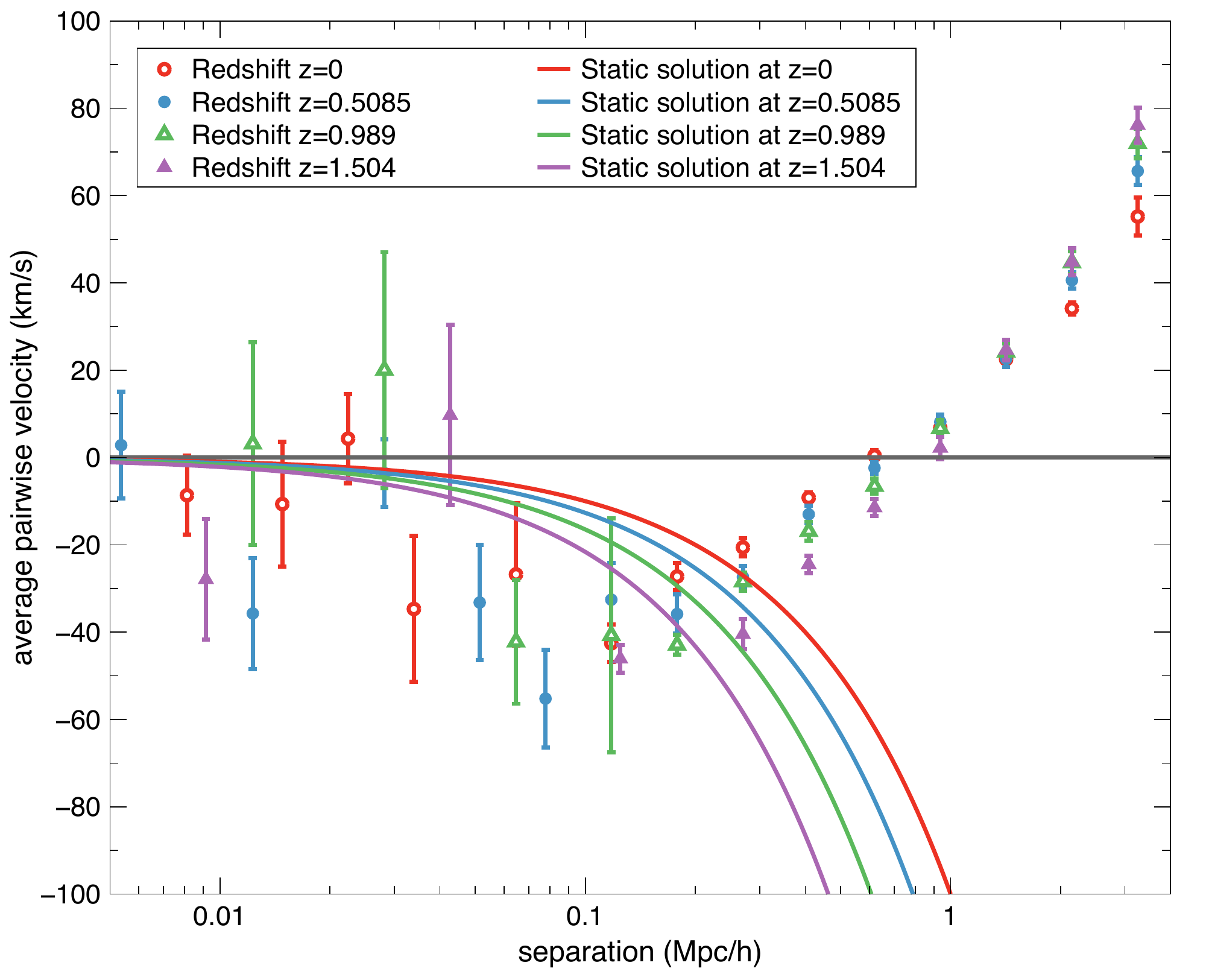}}
\subfloat[Mass $ \geqslant 4.3\times10^{11}$ M$_\odot/h$\, \;\;(\:$ \np \geqslant 500 \:)$] {\label{fig:np_100_iso}\includegraphics[width=0.5\textwidth]{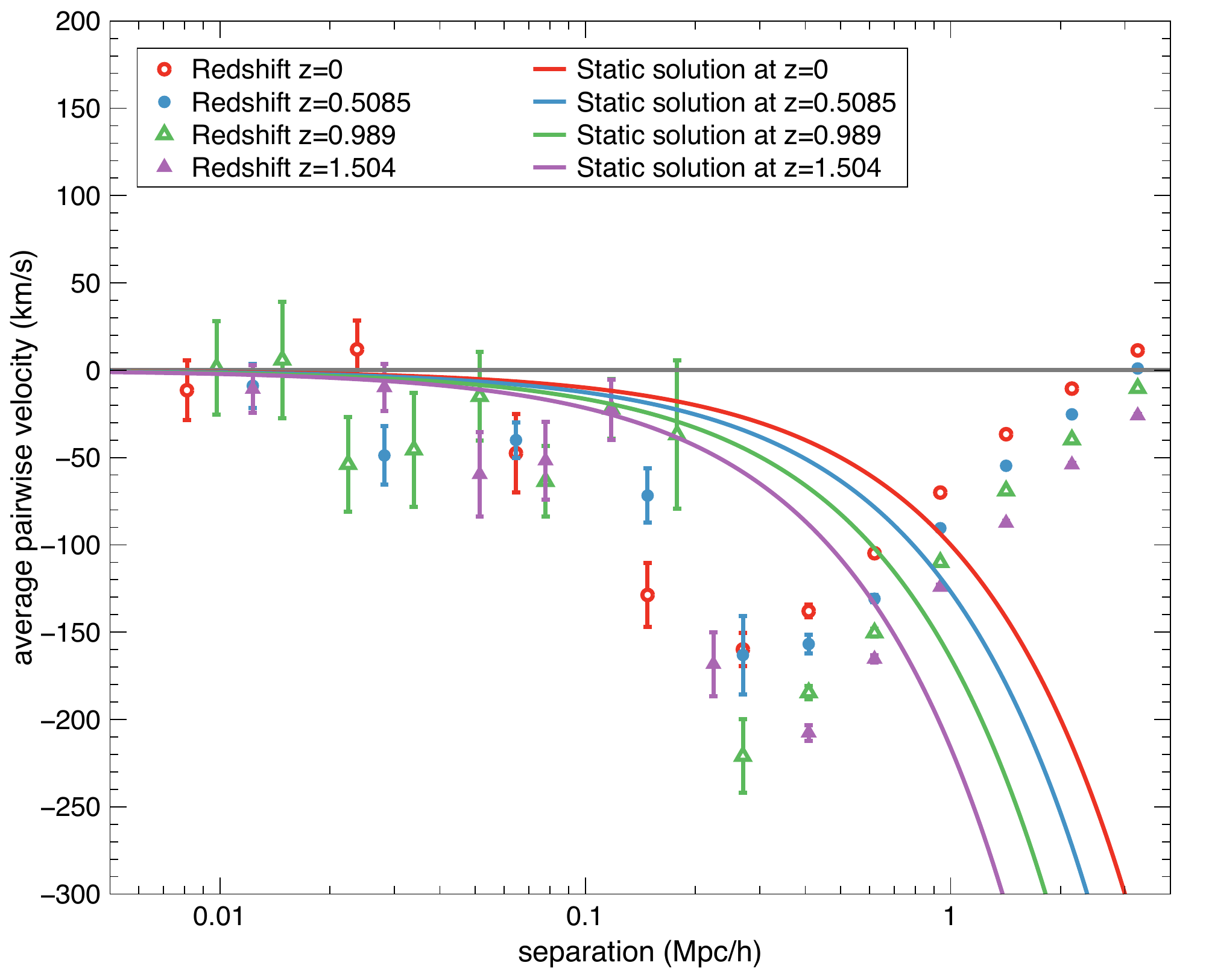}}\\
\subfloat[Mass $ \geqslant 8.6\times10^{11}$ M$_\odot/h$\, \;\;(\:$ \np \geqslant 1000 \:)$] {\label{fig:np_1000_iso}\includegraphics[width=0.5\textwidth]{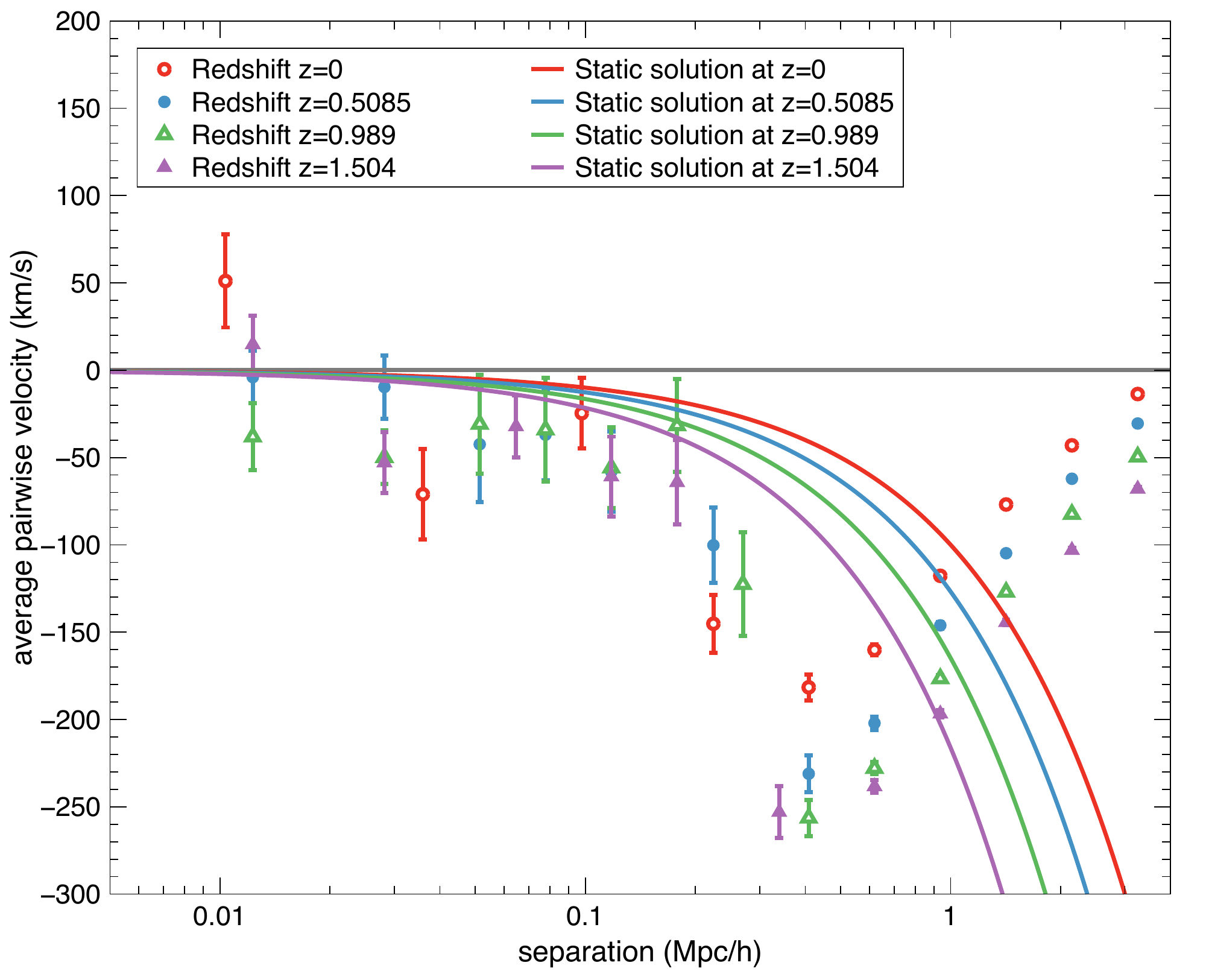}}
\subfloat[Mass $ \geqslant 4.3\times10^{12}$ M$_\odot/h$\, \;\;(\:$ \np \geqslant 5000 \:)$] {\label{fig:np_5000_iso}\includegraphics[width=0.5\textwidth]{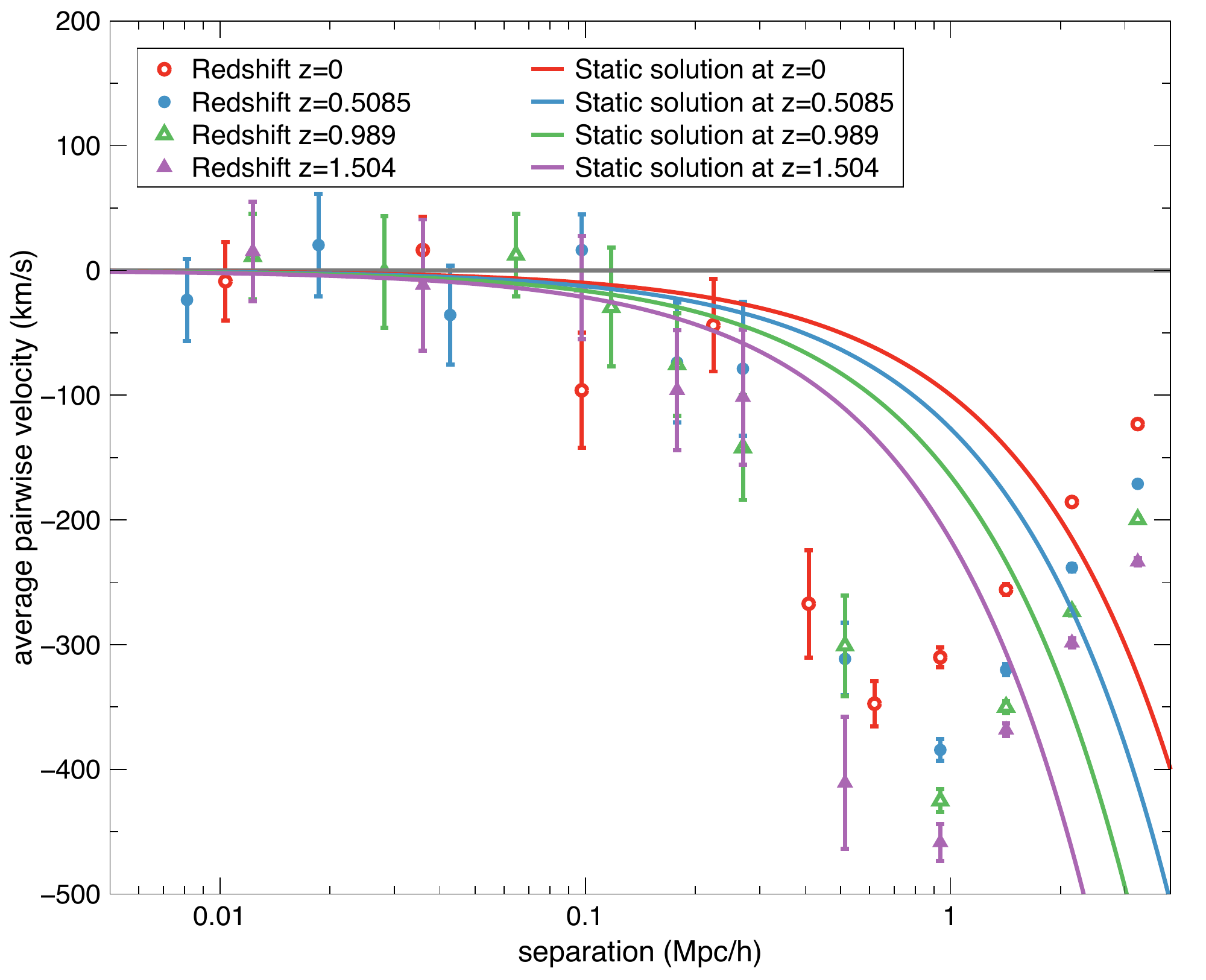}}
\caption{Variation of the average pairwise velocity of isolated galaxy pairs with redshift for different mass cuts (number of dark matter particles) as a function of separation. The isolation radius is taken to be \mpch{4} for each member of the pair. Note that panel \fref{fig:np_0_iso} is equivalent to the upper panel of \fref{fig:iso_redshift_variation}.}
\label{fig:masscut_variation_isolation4}
\end{figure*}

Having analyzed the dynamics of non-isolated pairs with varying subhalo mass, we now do the same for pairs isolated within a \mpch{4} radius. In \fref{fig:redshift_variation_isolation4} we show the average pairwise velocity $ v_{12} $ for different mass cuts, with the redshift varying form panel to panel. This is the same setup as in \fref{fig:redshift_variation_noiso}; note, however, that here we only plot separations up to \mpch{4}.

All curves in \fref{fig:redshift_variation_isolation4}, regardless of redshift, present the same features found in the uncut sample shown in \fref{fig:generic_isolated} and explained in \sref{sec:isolated_pairs}.  Namely, we see a virialized region on the smallest scales, an infalling regime on intermediate scales, and a roughly logarithmic growth due to the void effect on the largest separations analyzed.  The main difference from the uncut case is the separations at which these different regimes hold.  Most importantly, we can see that the logarithmic growth of $ v_{12} $ begins at larger separations for higher masses.  This is intuitive, since we expect the mutual attraction to be stronger in heavier pairs, thus overcoming the void effect even when the galaxies are closer to the edge of the void.

As a result of this stronger mutual attraction, the peculiar velocity contribution increases as we consider heavier pairs.  This means that when it comes to isolated galaxy pairs, low-mass pairs trace the cosmological expansion better than high-mass ones.  More quantitatively, the scale $ d_0 $ where the peculiar velocity vanishes is reached at larger separations for massive pairs.  At $ z=0.989 $, $ d_0 $ ranges from \mpch{0.8} in the uncut case to almost \mpch{4} for pairs with $ m>\msunh{4.3}{11} $.   For higher masses, $ v_{12} $ does not even cross the zero line.

\begin{figure*}[ht]
\subfloat[No isolation]{\label{fig:noiso_rmag}\includegraphics[width=0.5\textwidth]{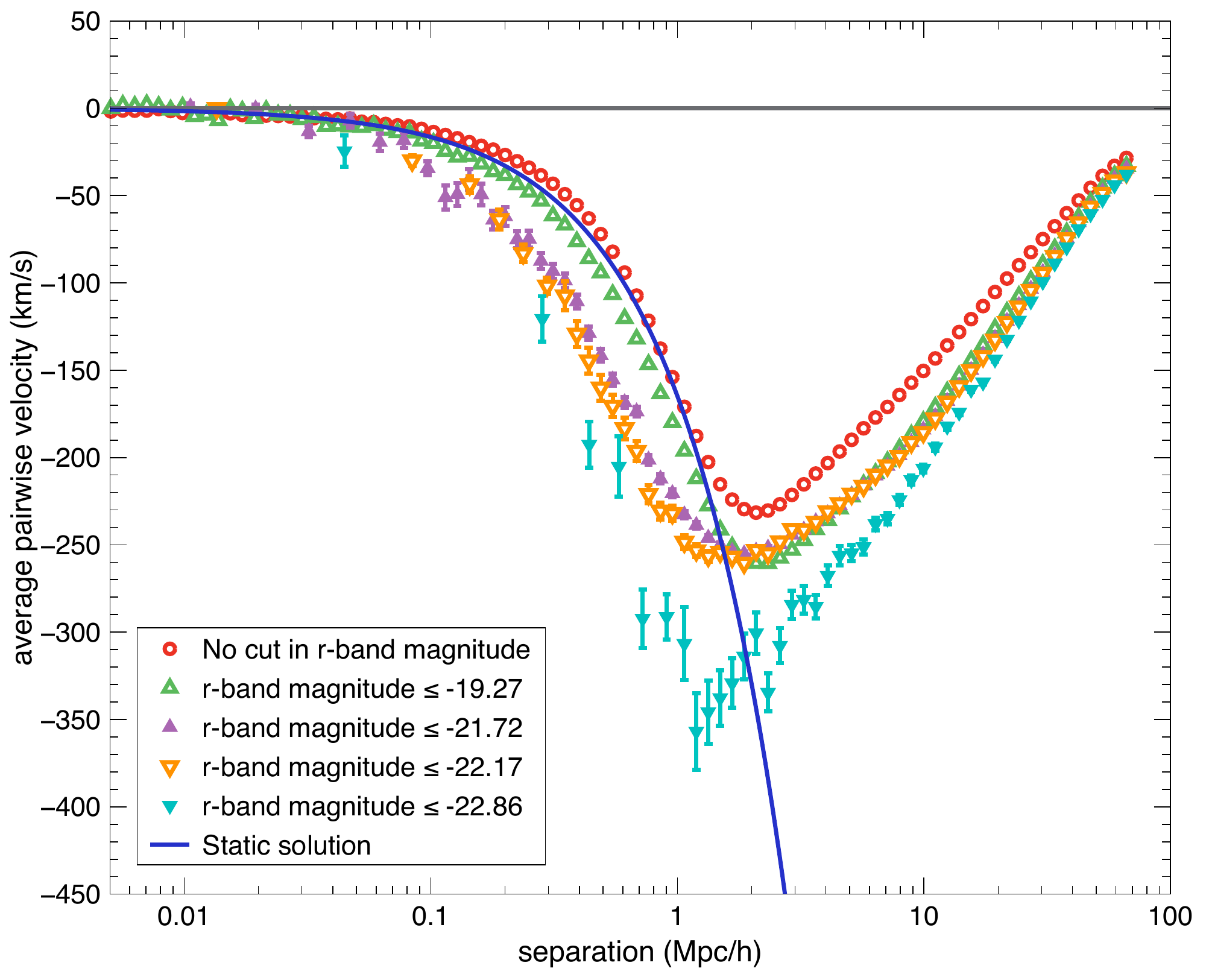}}
\subfloat[Isolation of \mpch{4}]{\label{fig:iso_rmag}\includegraphics[width=0.5\textwidth]{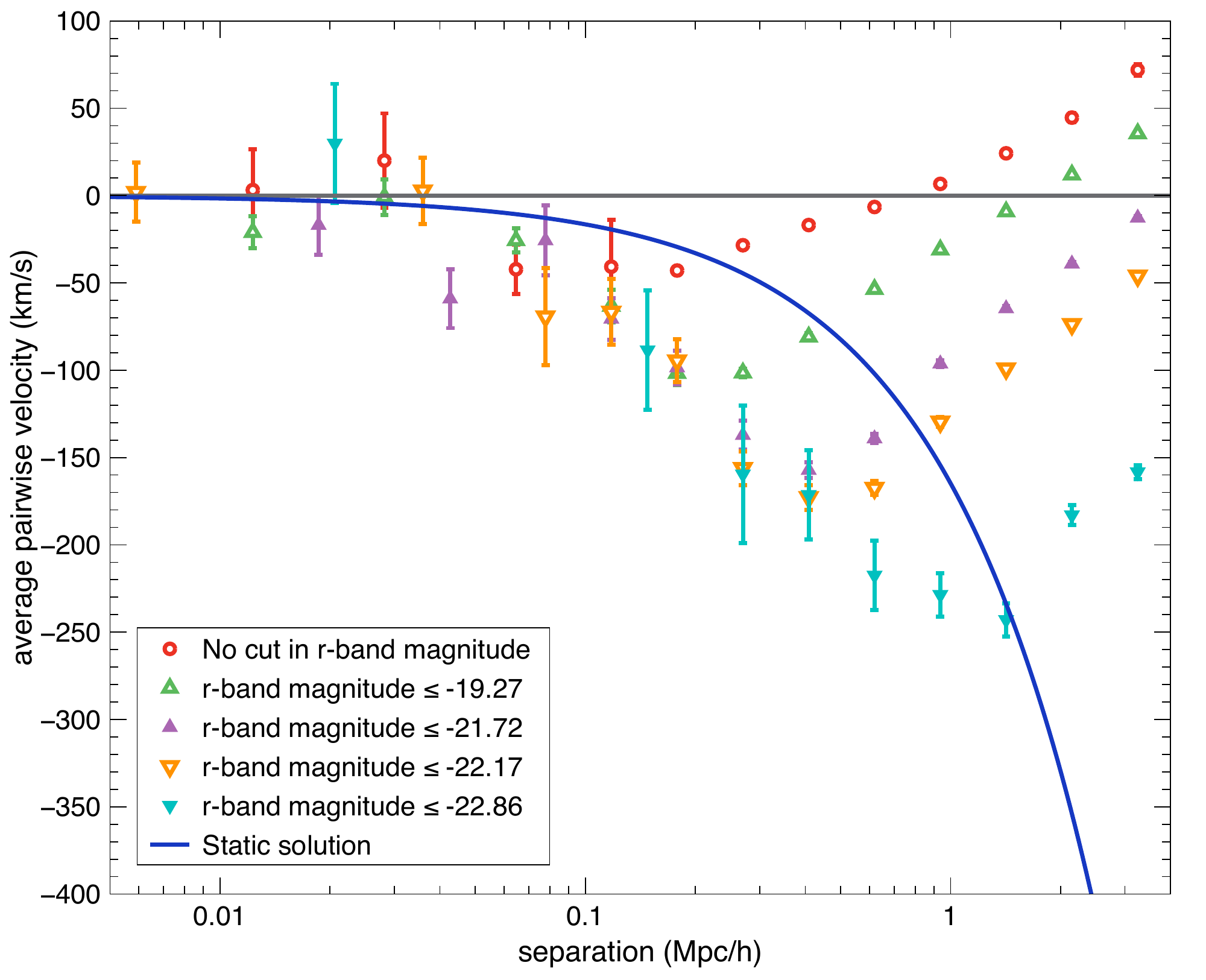}}
\caption{Average pairwise velocity of galaxy pairs at $z=0.989$ varying r-band magnitude.}
\label{fig:rmag_variation}
\end{figure*}

\begin{figure*}[ht]
\subfloat[No isolation]{\label{fig:noiso_stellarm}\includegraphics[width=0.5\textwidth]{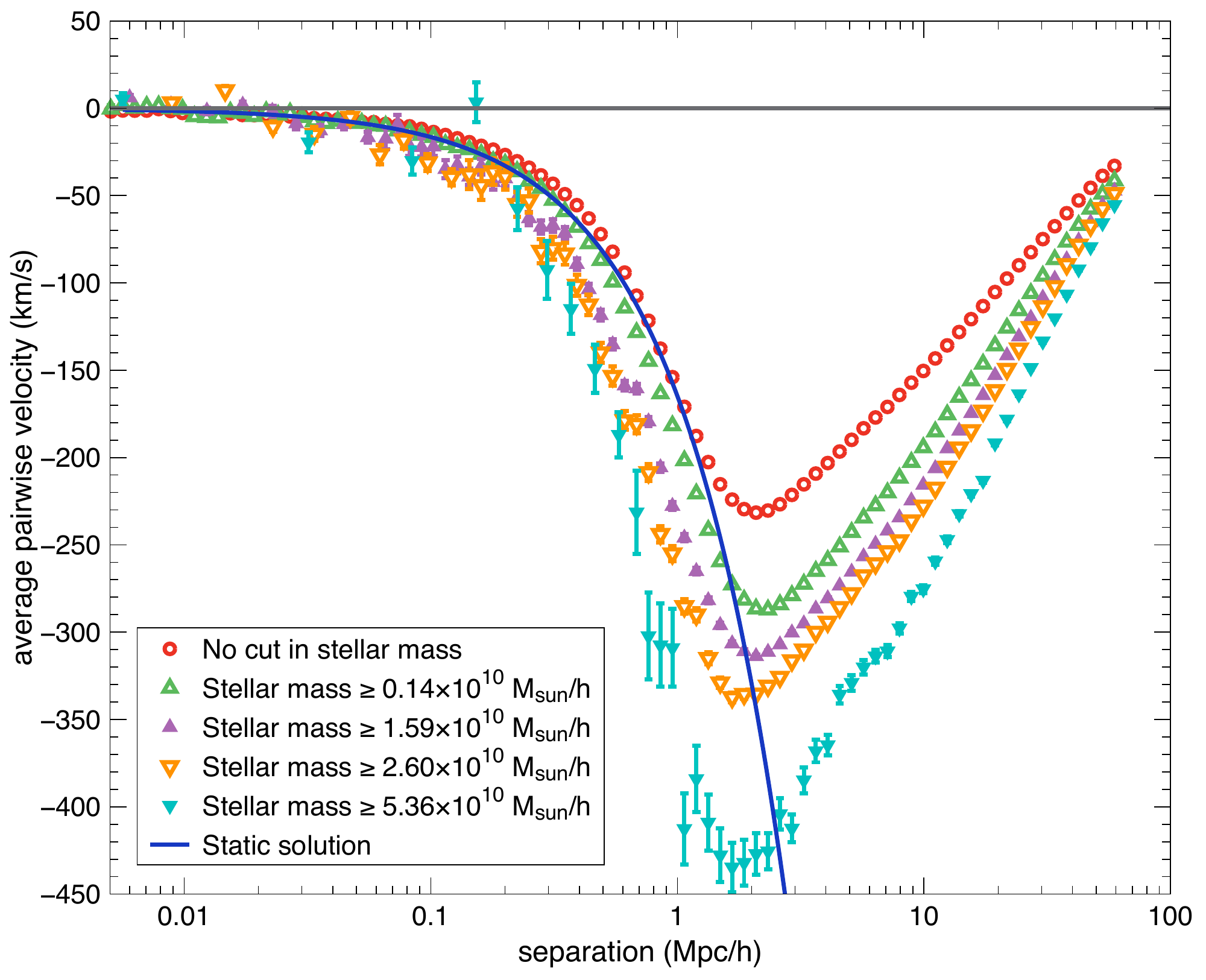}}
\subfloat[Isolation of \mpch{4}]{\label{fig:iso_stellarm}\includegraphics[width=0.5\textwidth]{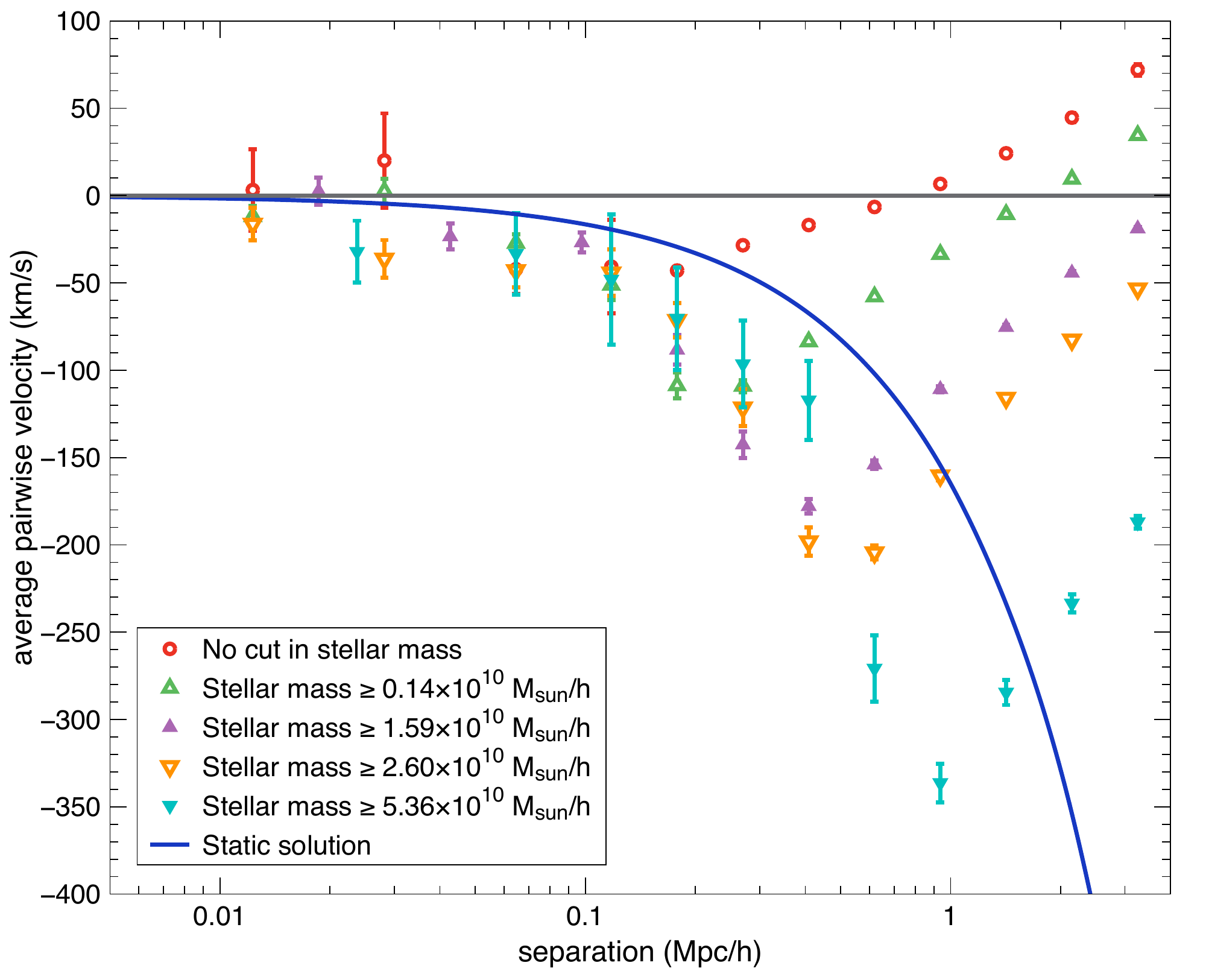}}
\caption{Average pairwise velocity of galaxy pairs  at $z=0.989$ varying stellar mass.}
\label{fig:stellar_mass_variation}
\end{figure*}

To illustrate the redshift dependence of the peculiar velocity in more detail, in \fref{fig:masscut_variation_isolation4} we plot $ v_{12} $ for a given mass-cut at four different redshifts, with the mass-cuts varying across the panels.  Increasing the mass-cut makes the redshift evolution of $ v_{12} $ more evident.  As a result, for $ m > \msunh{4.3}{11} $, we cannot identify a cosmological regime where the pairs are comoving and have a redshift independent peculiar velocity. Where ``independence" here means that the evolution is significantly less than the change in expansion rate.

A closer look at \fref{fig:redshift_variation_isolation4} shows a pattern in the different mass cuts. We notice that, as we increase the mass cut, the absolute value of the pairwise velocity increases and the minimum shifts to the right. This suggests that applying a mass dependent scaling to both separation and velocity may stack the curves. This is shown in \fref{fig:scaling_z1}, where we have have taken as an example the plot at redshift $z=0.989$ (\fref{fig:z_1_iso}) and scaled both the x and y axis by a factor $(M_{\text{ref}}/M)^{1/3}$ where $M_{\text{ref}}$ is the minimum mass of a subhalo: $M_{\text{ref}}=\sci{1.72}{10} \msun/h$. The physical motivation for this scaling is to have the same orbital period for all curves in the Keplerian regime \ie on small scales. Indeed we see on this plot that such scaling collapses the curves specially in the infalling and virialized regions.

\subsubsection{Magnitude \& stellar mass}

\begin{figure*}[ht]
\subfloat[No isolation]{\label{fig:noiso_font}\includegraphics[width=0.5\textwidth]{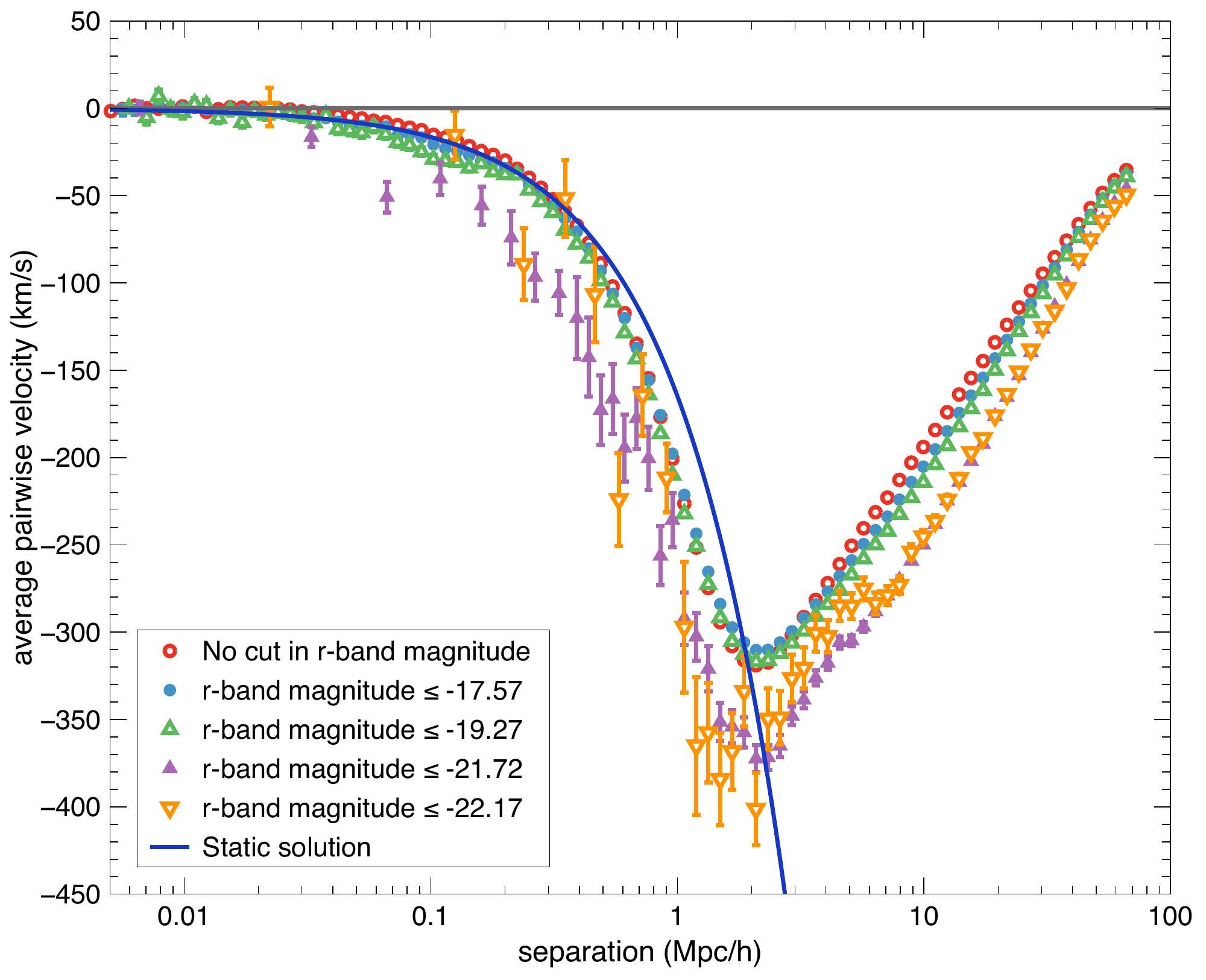}}
\subfloat[Isolation of \mpch{4}]{\label{fig:iso_font}\includegraphics[width=0.5\textwidth]{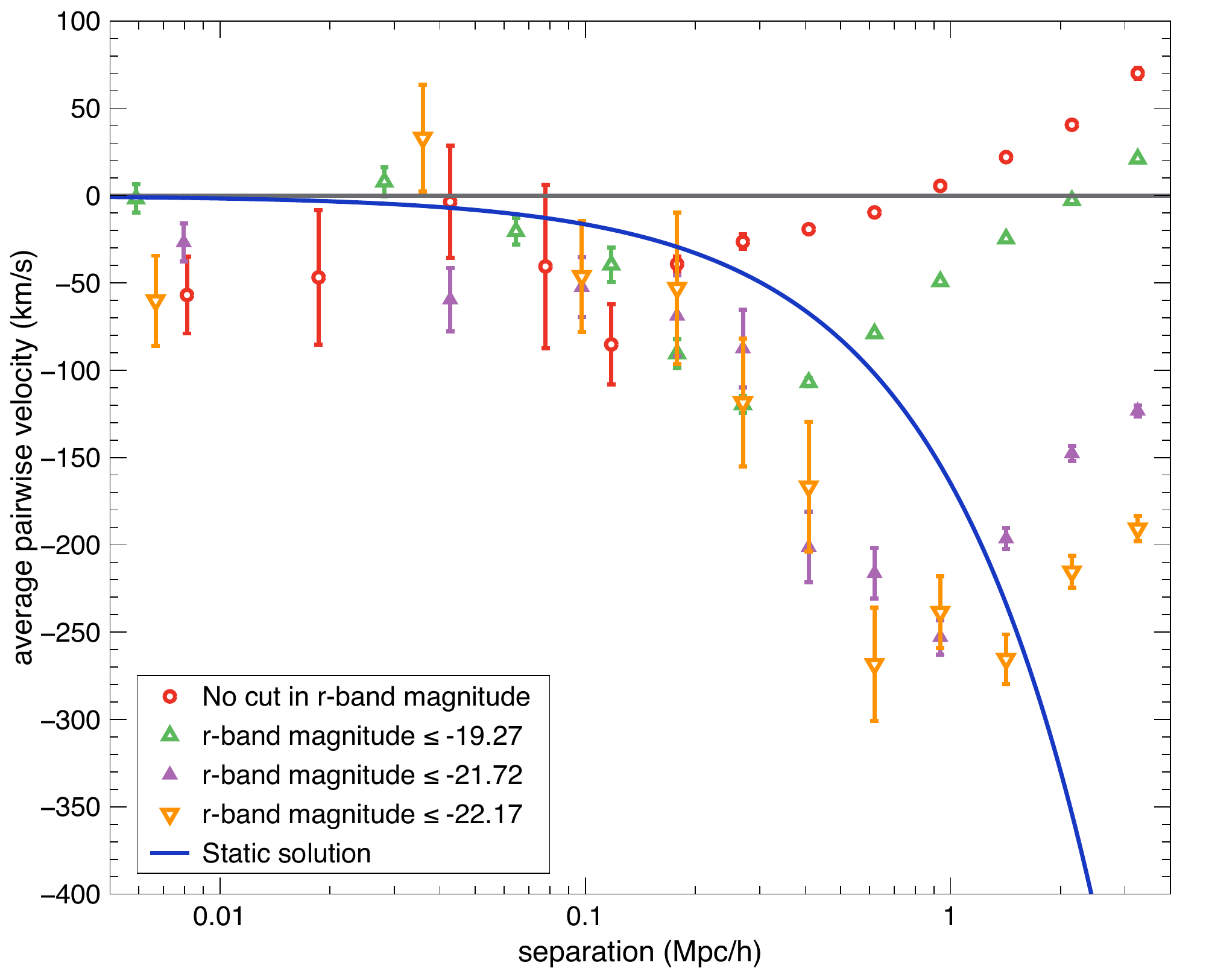}}
\caption{Average pairwise velocity of galaxy pairs for the \citet{font:2008a} semi-analytic model at $ z=0.989 $.  This figure should be compared with \fref{fig:rmag_variation}, where we plotted the same curves obtained for the \citet{guo:2011a} model.}
\label{fig:rmag_font}
\end{figure*}

We now make a more direct link with observations and study the dependence of $ v_{12} $ on $ r $-band absolute magnitude and stellar mass.

In the left panel of \fref{fig:rmag_variation}, we show the average pairwise velocity $ v_{12} $ of all galaxy pairs at $ z = 0.989 $ for the \rmag cuts given in \tref{tab:cuts}.  Although the curves retain their qualitative shape, there are two major differences with respect to the mass-cut sample in \fref{fig:z_1}.  Firstly, the \rmag selected pairs have smaller average velocities.  Secondly, the velocity minima are all approximately aligned at the same scale of $r_\text{min}\sim$ \mpch{2}, while for the subhalo mass cuts the different velocity curves have their minima at different separations. These two differences are also seen when we apply the cuts in stellar mass (\fref{fig:noiso_stellarm}).

The velocity differences can be explained by the fact that, although the subsamples chosen based on limits in $ r $-band magnitude and stellar mass preserve the number density of galaxies selected, these are not the same galaxies as the ones selected by the subhalo mass cuts. In particular, most massive galaxies do not necessarily coincide with the most luminous ones. In general, dark matter haloes trace the velocity of galaxies more directly than stellar mass or \rband magnitude. Cuts based on stellar mass or luminosity add an additional dispersion, affecting the position of the minima with respect to subhalo mass cuts.

The right panels of \fref{fig:rmag_variation} and \fref{fig:stellar_mass_variation} show the average pairwise velocity $ v_{12} $ for pairs with an isolation radius of \mpch{4} for cuts in \rband magnitude and stellar mass respectively. These plots should be compared with the corresponding cuts in subhalo mass at redshift $z=0.989$ (\fref{fig:z_1_iso}). Even though we again appreciate that the pairwise velocities in the \rmag and stellar mass cut plots are smaller, the general dynamics shown on the plots are the same. It is worth noting that the almost comoving regime mentioned in \sref{sec:cosmological_implications} for each curve remains unchanged both for the \rband magnitude and the stellar mass cuts. It is clear that the effects of galaxy selection (be it subhalo mass, stellar mass or \rband magnitude) play an important role in the behavior of pairwise velocities for isolated galaxy pairs.

\section{Comparison of two catalogs} \label{sec:comparison_catalogs}
The results presented in the previous sections were based on the semi-analytic model of \citet{guo:2011a}. To check the robustness of these results, we also compute the average peculiar velocities $ v_{12} $ for the semi-analytic model in \citet{font:2008a}. This catalog is an improvement over the one presented in \citet{bower:2006a} to better match the colors of satellite galaxies observed in the SDSS sample. In order to do this, the main modification introduced in \citet{font:2008a} is the stripping of hot gaseous haloes of satellite galaxies into the GALFORM semi-analytic model for galaxy formation, while \citet{guo:2011a} concentrate more on the independence of satellite galaxies from the FOF group. Both semi-analytic models have similar galaxy luminosity functions that fit the data well.

In \fref{fig:rmag_font} we show the average pairwise velocity $ v_{12} $ as a function of separation for all the galaxy pairs (left panel) and for isolated pairs with an isolation radius of \mpch{4} (right panel).  Each curve corresponds to one of the \rmag cuts in \tref{tab:cuts}, and should be compared with the matching curve for the \citet{guo:2011a} catalog in \fref{fig:rmag_variation}.   Note that we omitted to plot $ v_{12} $ for our most stringent cut of $ \rmag < -22.86 $ because of poor statistics.  In general, we found that \citet{font:2008a} has significantly less bright galaxies with $ \rmag < -22.17 $ than \citet{guo:2011a}, as can be seen by the large error bars in \fref{fig:rmag_font}. 

A direct comparison between Figures \ref{fig:rmag_font} and \ref{fig:rmag_variation} shows that galaxy pairs have very similar dynamics regardless of the semi-analytic model used.   Not only do we see almost the same $ v_{12} $ range, but also the almost comoving regime introduced in \sref{sec:cosmological_implications} is found approximately in the same range.  Such findings suggest that the parameters of the semi-analytic model used do not significantly affect the average pairwise velocities of galaxies.

\section{Conclusions}  \label{sec:conclusions}
We have investigated the pairwise velocities of galaxies with a view towards using them as cosmological tracers by means of an AP style test, as recently proposed by \citet{marinoni:2010a}.  We have analyzed the dynamics of such objects within the semi-analytic models of \citet{guo:2011a} and \citet{font:2008a} applied to the Millennium simulation \cite{springel:2005a}, and studied the dependence of their relative velocity on local density, redshift, mass of the hosting subhaloes, \rband magnitude and stellar mass.

We have first analyzed the dynamics of all galaxy pairs at redshift $ z=0 $ (see \fref{fig:generic_all_pairs}).  We have found that, on scales $ d>\mpch{10} $, the peculiar velocity is correctly predicted by linear theory \cite{fisher:1995a, reid:2011a}.  On the other hand, for separations $ d<\mpch{3} $, the pairs are decoupled from the Hubble flow and close to the static solution.  We argue that pairs in this regime cannot be used as cosmological tracers (see Appendix \ref{app:bound_systems}).  

Being interested in investigating the claims by \citet{marinoni:2010a}, we have studied the dynamics of galaxy pairs that are isolated within a radius of $ \mpch{4} $.  At $ z=0 $, isolated galaxy pairs are almost comoving already for separations of $ 0.4<d<\mpch{4} $ and only need up to 20\% RSD correction (see \fref{fig:generic_isolated}).   By analyzing redshift slices up to $ z=1.504 $, we have found that the peculiar velocities are only weakly dependent on the cosmological expansion ($ < 10\% $ variation) for separations of $ 1<d<\mpch{4} $ (see \fref{fig:iso_redshift_variation}).  Since expansion is the main property characterizing a cosmological model, we might assume that in this regime the dynamics of isolated pairs are independent of the underlying cosmology.  Hence, we argue that isolated pairs in this regime could possibly be used as cosmological tracers with minimal RSD corrections.  

Imposing an isolation criterion of \mpch{4}, as done in Ref.\ \cite{marinoni:2010a}, greatly reduces the number of pairs (see \fref{fig:number_pairs}).  We have found that one can drastically increase the statistics while keeping the RSD corrections small by either reducing the isolation radius to \mpch{2} or allowing up to $ 10 $ galaxies to be neighbors of the pair.  When dealing with observations, these adjustments may be helpful to reduce possibly large statistical uncertainties.

As galaxy surveys are flux limited, we have studied the feasibility of a measurement by varying the following properties of galaxy pairs: mass of the subhaloes that host the galaxies, \rband absolute magnitude in the rest-frame, and stellar mass.   Low-mass pairs appear to be the best cosmological tracers, as RSD corrections increase with mass.  More precisely, a nearly comoving regime is reached in our analysis only for subhalo masses of $ m \lesssim \msunh{4.3}{11} $, corresponding to \rband magnitudes of $ \rmag \gtrsim -21.27 $ and stellar masses of $ \smass \lesssim \msunh{1.59}{10} $ (see, respectively, Figures \ref{fig:redshift_variation_isolation4}, \ref{fig:rmag_variation} and \ref{fig:stellar_mass_variation}).  We have also found that the peculiar velocities of galaxy pairs becomes more redshift-dependent as we increase the subhalo mass (see \fref{fig:iso_redshift_variation}).  Therefore, we suggest that isolated pairs may not be adequate as cosmological tracers if their mass or luminosity is above the given thresholds.

Marinoni and Buzzi \cite{marinoni:2010a} selected isolated galaxy pairs from the DEEP2 galaxy survey \cite{davis:2003a, newman:2012a} with comoving transverse separation $ \rperp $ in the range $ \unit[20]{kpc/h} $ -- $ \mpch{0.7} $.  DEEP2 galaxies are known to reside in dark matter haloes of approximately $ \unit[10^{12}]{\msun/h} $ \cite{newman:2012a, conroy:2005a}.  The results of our analysis imply that such galaxy pairs do require corrections for evolution and cosmology dependent RSD component, which is significant with respect to the evolution being measured. 

Indeed, the primary concern for observational studies is the extent to which RSD ``corrections" need to be modeled. Cosmological measurements from Baryon Acoustic Oscillation and RSD measurements on large-scales are reaching a precision at the 2--5\% level \cite{anderson:2012a,reid:2012a,padmanabhan:2012a}, and it is therefore reasonable to suppose that this is also the level at which we need to understand RSD corrections in order to make a useful contribution to the field from small-scale measurements. We have investigated whether selection based on local density can reduce the modeling burden, and find that low-mass, isolated galaxy pairs are preferred. However, even for these galaxies, the corrections depend on sample properties, and would need to be recalculated for each cosmological model to be tested: the only currently available way to do this is via numerical simulations. We conclude that observations of close-pairs of galaxies do show promise for AP-style cosmological measurements, particularly for low mass, isolated galaxies. However, it is likely that modeling limitations will continue to be the limiting factor for the foreseeable future.

\section{Acknowledgments}
ABB acknowledges financial support from the Spanish Ministry 
of Education (ME) within the FPU grant program with ref.\ number AP2008-02679. 
She also thanks the Institute of Cosmology and Gravitation for hospitality 
during her stay from September 2011 to January 2012. WJP thanks the European Research Council and the UK Science and Technology Facilities Research Council for financial support. The authors would like to thank David Bacon, Marco Bruni, Phil Bull, Diego Capozzi, Chris Clarkson, Timothy Clifton, Robert Crittenden, Chris D'Andrea, Sean February, Pedro G.\ Ferreira, Juan Garc\'ia-Bellido, Alan Heavens, Elise Jennings, Marc Manera, Francesco Pace, John A.\ Peacock, Mat Pieri and Lado Samushia for useful discussions. We would also like to thank the referee for useful suggestions.

\bibliographystyle{apsrev4-1}
\bibliography{my_bibliography_short}

\begin{appendix}

\section{Observing bound systems}\label{app:bound_systems}

\begin{figure}
  \begin{center}
    \includegraphics[width=0.45\textwidth]{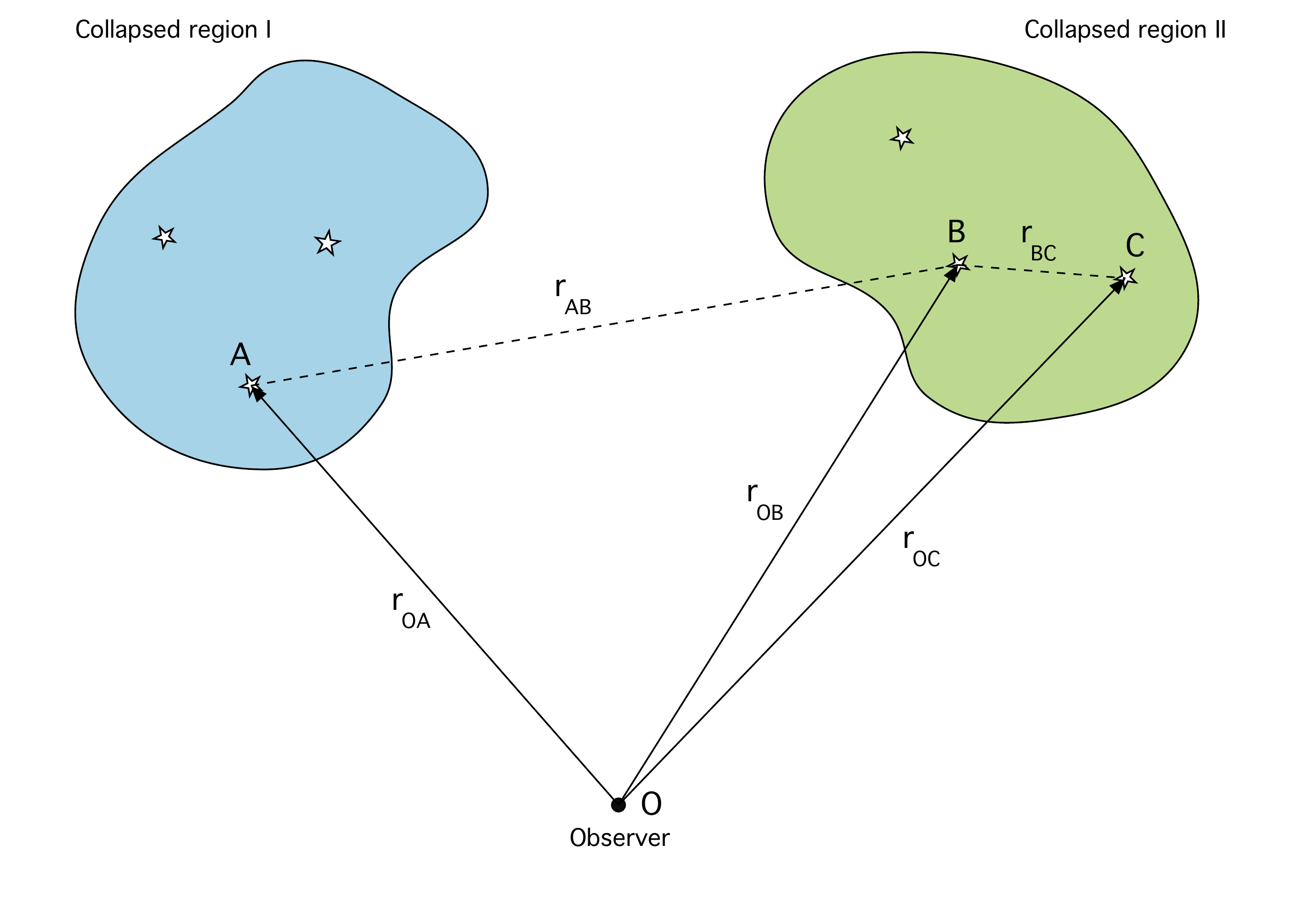}
  \end{center}
\vspace{-0.3cm}
  \caption{This figure shows schematically two collapsed
    regions. Objects $A$ and $B$, which are not gravitationally bound,
    will have different cosmological redshifts given by $z_{A,B}=H_0
    r^{A,B}_{\text{com}}+v^{A,B}_\|/c$, while gravitationally bound
    objects $B$ and $C$, shown in the second collapsed region, will have a line of
    sight component of their velocity that exactly cancels the
    expansion so that particles within that region will all have the
    same cosmological redshift.}
\label{fig:patches}
\end{figure}

Consider applying the AP effect for bound systems, such as shown by
the triangle $OBC$ in Fig.~\ref{fig:patches}. Here we cannot relate
$\Delta z$ to the proper distance using the Hubble parameter, or if
we force this, we have to consider a peculiar velocity that cancels
the expansion. We can write the observed redshift width
\begin{equation}
  \Delta z^0=\Delta z+\frac{dv_\parallel}{c}(1+z).
\end{equation}
If $v_\parallel$ is the orbital motion only $v_\parallel=v_{\rm orb}$, then
$\Delta z=0$ as photons from both galaxies are subject to the same
cosmological expansion. 

We can calculate a variable with the units of distance as in \cite{marinoni:2010a} 
\begin{equation}
\Delta x^0=\Delta x + \frac{dv_\parallel}{H(z)}(1+z).
\end{equation}
But using the arguments above $\Delta x=0$ and, as $dv_\parallel$ is
independent of $H(z)$, using $H(z)$ to translate to distance does not
provide any extra cosmological information. Hence any information,
even if from apparent orientation of pairs, is independent of $H(z)$.

We now consider bound systems that have broken free from cosmological expansion. In general, the Hubble expansion velocity can be defined for any pair
of particles in the Universe. With this definition, a static body has
peculiar velocities that oppose and balance the expansion. To see this
more clearly, consider an N-body simulation with a comoving coordinate
system. In this coordinate system, a body that has a constant proper size would
appear to be collapsing. In interpreting
this system from an N-body simulation, one might consider this
infalling velocity as a peculiar velocity, although in effect this is
simply balancing the cosmological expansion. One therefore sees that
there is a general interplay between the expansion rate and the
peculiar velocities, which must be included in any interpretation of
data.

Interpreting this in terms of local curvature, Fig.~\ref{fig:patches} shows two collapsed regions being observed. In
the standard interpretation, objects $B$ and $C$, which are in a
collapsed system, have peculiar velocities that cancel any
cosmological redshift between them. The infall peculiar
velocity must therefore be 
\begin{equation}
v_{pec}=-H(z)r_{BC}.
\end{equation}
A light ray sent from $B$ to $C$, will experience a Doppler shift due to
the motion of the objects towards each other in addition to the
cosmological redshift. Assuming that the light ray is emitted at a
wavelength $\lambda_{em}$, the Doppler shift changes this wavelength
to $\lambda_{dop}$ and the observed wavelength at $B$
$\lambda_{obs}$. The change in the wavelength due to the Doppler
shift, assuming velocities much smaller than the speed of light, is
\begin{equation}\label{eq:dop}
  \lambda_{dop}=\frac{\lambda_{em}}{\left(1+\frac{H(z)d}{c}\right)}.
\end{equation}
Due to the cosmological redshifting, the light ray is then observed at a wavelength of 
\begin{equation}
  \lambda_{obs}=\left(1+\frac{H(z)d}{c}\right)\lambda_{dop},
\end{equation}
where we can substitute \eqref{eq:dop} and obtain
\begin{equation}
  \lambda_{obs}=\lambda_{em}.
\end{equation}
This shows that if one treats the redshift difference between
two objects as including a cosmological expansion component, one cannot assume that the peculiar velocity is independent of cosmology: for bound systems their combined effect is zero. In an
alternative and equally valid interpretation, $B$ and $C$ live in a
flat Minkowski space-time, which does not lead to a cosmological
redshift due to cosmological expansion. Photons only start to
experience a cosmological redshift once they are free from the bound
system, and subject to cosmological expansion: photons from $B$ and
$C$ traveling to $O$ experience the same cosmological redshift. The line of sight radial velocity distribution is independent of $H(z)$ due to decoupling from cosmological expansion and hence, whatever the true
expansion rate $H(z)$, we should expect the same velocities from
isolated bound systems, which are simply all behaving as if they were
in Minkowski space-time.

\end{appendix}

\end{document}